\documentclass[aps,prd,10pt,superscriptaddress,twocolumn,nolongbibliography,shortbibliography,showkeys,serif,nofootinbib]{revtex4-2}
\usepackage{graphicx}
\usepackage{xcolor}
\usepackage{hyperref}
\usepackage{amsfonts, amsmath, amssymb}
\usepackage{lmodern}
\usepackage{dcolumn}
\usepackage{multirow}
\usepackage{enumitem}
\usepackage{booktabs}
\usepackage{enumitem}
\usepackage{subcaption}
\usepackage[sort&compress]{natbib}
\usepackage{soul}

\usepackage[dvipsnames]{xcolor}

\newcommand{\orcid}[1]{\href{https://orcid.org/#1}{\includegraphics[width=8pt]{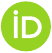}}}

\hypersetup{colorlinks=true, linkcolor=blue, urlcolor=blue, citecolor=blue}

\graphicspath{{figs/}} 

\begin{document}

\title{Exploring the Impact of Systematic Bias in Type Ia Supernova Cosmology Across Diverse Dark Energy Parametrizations}

\author{Drishti Sharma \orcid{0009-0004-4664-0860}}
\email{drishti2306519@st.jmi.ac.in}
\affiliation{Department of Physics, Jamia Millia Islamia, New Delhi-110025, India}%

\author{Purba Mukherjee \orcid{0000-0002-2701-5654}}
\email{pdf.pmukherjee@jmi.ac.in}
\affiliation{Centre for Theoretical Physics, Jamia Millia Islamia, New Delhi-110025, India}%

\author{Anjan A Sen \orcid{0000-0001-9615-4909}}%
\email{aasen@jmi.ac.in}
\affiliation{Centre for Theoretical Physics, Jamia Millia Islamia, New Delhi-110025, India}%

\author{Suhail Dhawan \orcid{0000-0002-2376-6979}}%
\email{s.dhawan@bham.ac.uk}
\affiliation{School of Physics \& Astronomy and Institute of Gravitational Wave Astronomy, University of Birmingham, UK}%


\begin{abstract}
We investigate the impact of instrumental and astrophysical systematics on dark energy (DE) constraints from Type Ia supernova (SN-Ia) observations. Using simulated datasets consistent with current SN-Ia measurements, we examine how photometric calibration, intergalactic dust, progenitor evolution in luminosity and light-curve stretch, intrinsic color scatter, {and matter density mismatch} affect the inferred DE equation of state (EoS) parameters $(w_0,w_a)$. We test the Generalised Scale Factor (GEN) parametrization against three time-evolving DE models: Chevallier-Polarski-Linder (CPL), Jassal-Bagla-Padmanabhan (JBP), and Logarithmic (LOG). Calibration and progenitor-related effects emerge as the dominant sources of bias. {In particular, a calibration offset of $\Delta M_B=0.02$ can shift the inferred parameters by up to $\Delta w_0 \simeq -0.12$ and $\Delta w_a \simeq +0.60$ in JBP, while the corresponding shift in GEN is much smaller, with $\Delta w_0 \simeq -0.02$ and $\Delta w_a \simeq -0.04$. Progenitor-stretch evolution also induces substantial shifts, whereas intergalactic dust and color-scatter systematics produce only minor deviations for the fiducial amplitudes adopted here. Overall, JBP is the most sensitive to injected systematics, CPL and LOG show intermediate sensitivity, and GEN remains the most stable. We also quantify the deviation from the fiducial $\Lambda$CDM ($w_0=-1,\, w_a=0$) for the injected-systematic cases, and find that the second set of systematic injections supports the same qualitative hierarchy.} These results highlight the need for sub-percent calibration precision and improved astrophysical modelling for robust DE inference from present and future SN-Ia cosmology experiments. More broadly, our results motivate model-independent tests of late-time physics, with phenomenological $(w_0,w_a)$ parametrizations used as summary statistics.
\end{abstract}

\keywords{cosmology, supernovae, distance modulus, dark energy, systematics}

\maketitle


\section{Introduction \label{sec:intro}}

Over the past few decades, cosmology has advanced dramatically, driven by high-precision measurements from the cosmic microwave background (CMB) \cite{Planck:2018vyg, Tristram:2023haj, ACT:2025fju}, Type Ia supernovae (SN-Ia) \cite{Pan-STARRS1:2017jku, Brout2022}, baryon acoustic oscillations (BAO) and large-scale structure (LSS) \cite{BOSS:2016wmc, eBOSS:2020yzd, DESI:2024jxi} surveys. Together, these observations form the basis of the standard cosmological model, $\Lambda$CDM, which describes a spatially flat universe composed primarily of cold dark matter (CDM) and dark energy (DE). Although the cosmological constant ($\Lambda$) serves as the simplest form of {DE} and successfully explains the observed late-time acceleration, its physical interpretation remains unresolved \cite{Weinberg:1988cp, Peebles:2002gy, Padmanabhan:2002ji}. Hence, understanding the physical origin of this `dark energy' is a fundamental question in modern cosmology.

The discovery of accelerated expansion of the universe in late times, inferred from the luminosity–distance relation of SN-Ia, remains one of the most significant results in cosmology \cite{SupernovaSearchTeam:1998fmf, SupernovaCosmologyProject:1998vns}. Since then, SN-Ia have been established as a cornerstone of empirical {DE} studies, allowing measurements of the {DE} equation of state (EoS) parameter ($w = {p_{\rm DE}}/{\rho_{\rm DE}}$) \cite{Goobar_2011, Suzuki_2012, SDSS:2014iwm}. The most widely used time-evolving {DE} model is the Chevallier-Polarski-Linder (CPL) \cite{Chevallier:2000qy, Linder:2002et} parametrization,  given by $w(a) = w_0 + w_a (1 - a)$, where $w_0$ denotes the present-day value of the EoS, $w_a$ describes its evolution with the scale factor $a$. When combined with CMB anisotropies and BAO distance measurements, SN-Ia provide powerful constraints on $w_0$ and $w_a$, which for $w_0=-1$ and $w_a=0$ reduce to the concordance $\Lambda$CDM scenario \cite{Brout2022, DES:2024jxu, Rubin:2023jdq}. Despite its successes, $\Lambda$CDM faces several challenges: local measurements of the Hubble constant differ from CMB-inferred values by $>5\sigma$ \cite{H0DN:2025lyy}, the amplitude of matter fluctuations exhibits a mild tension with predictions from the CMB \cite{Perivolaropoulos:2021jda, Riess:2022oxy, DiValentino:2024yew, Efstathiou:2024dvn, CosmoVerseNetwork:2025alb}. Observations of unexpectedly massive galaxies by the James Webb Space Telescope (JWST) at high redshift also challenge the concordance framework \cite{Labbe:2022ahb, Boylan-Kolchin:2022kae}. 

Recent high-precision galaxy redshift surveys, such as Dark Energy Spectroscopic Instrument (DESI), allow accurate measurements of the {BAO} scale over $0.1 < z < 3.5$ \cite{DESI:2025zpo}. When combined with CMB and SN-Ia data, these observations hint at a time-varying {DE} component at the $\sim 2.8-4.2\sigma$ level \cite{DESI:2025zgx} (depending on the SN-Ia sample and calibration method), motivating the exploration of dynamical {DE} beyond the canonical $\Lambda$CDM paradigm \cite{DESI:2025fii, DESI:2025wyn}. Recent results by Mukherjee \& Sen \cite{Mukherjee:2025ytj} identified seven characteristic redshifts, where the expansion history reconstructed using SN-Ia and BAO (anchored to Planck-2018 $\Lambda$CDM physics) shows $>3\sigma$ deviations from Planck $\Lambda$CDM predictions for $z<1$. As a follow-up, Choudhury, Mukherjee \& Sen \cite{GuptaChoudhury:2025uff} examined the robustness of these characteristic redshifts and found them to remain stable under diverse {DE} parametrizations. Therefore, such discrepancies, if not for new physics beyond $\Lambda$, underscore the necessity of rigorously assessing systematic effects that can bias the cosmological parameter inference. 

The reliability of supernova cosmology depends on accurate relative distance measurements across redshift. Redshift-dependent variations in supernova populations, calibration, or host-galaxy environments can introduce biases that mimic or obscure true cosmological signals \cite{Nordin:2010ti}. Low-redshift SN-Ia serve as the distance anchor for the Hubble diagram \cite{Riess:2021jrx, Brout2022, Riess:2024vfa}, but historically, they were observed with heterogeneous instruments, leading to significant calibration uncertainties \cite{Brout:2021mpj, Efstathiou:2024dvn}. Modern wide-field surveys, such as the Zwicky Transient Facility (ZTF), now provide a uniformly observed and well-calibrated low-redshift sample, while contemporary high-redshift surveys offer greatly improved statistical precision \cite{Graham:2019qsw, Dhawan_2021, Rigault2024}. The combination of these datasets is expected to enable a thorough assessment of systematic effects and their effect on {DE} measurements.
In this work, we investigate the impact of instrumental and astrophysical systematics on {DE} constraints derived from {SN-Ia}, inspired by previous studies by \cite{Nordin:2010ti} and \cite{Dhawan:2024gqy}. We consider a range of effects, including photometric calibration errors \cite{Brout:2021mpj}, intergalactic dust \cite{Goobar2018}, progenitor evolution affecting luminosity \cite{Dhawan2024Encore} and light-curve stretch \cite{Ginolin_2025}, as well as increased intrinsic scatter in color corrections \cite{Popovic2023}. Using simulated datasets consistent with current SN-Ia observations, we evaluate how these systematics influence the inferred {DE} parameters. We test the Generalised Scale Factor (GEN) \cite{Sen:2001xu, Mukhopadhyay2024} evolution, analysing its susceptibility to the injected systematics, and compare the results with three time-evolving {DE} models, namely {Chevallier–Polarski–Linder (CPL)}, Jassal–Bagla–Padmanabhan (JBP) \cite{Jassal2005}, and Logarithmic (LOG) \cite{Efstathiou1999} parametrizations. By quantifying the sensitivity of $(w_0, w_a)$ to each source of uncertainty, we identify which systematics are most likely to mimic the hints of evolving {DE} observed in recent combined SN-Ia, CMB, and DESI datasets, corresponding to deviations of $\sim 3\sigma$ from the concordance $\Lambda$CDM values of $w_0 = -1$ and $w_a = 0$ \cite{DESI:2025zgx}. This analysis establishes benchmarks for understanding how unaccounted systematic effects can bias cosmological inference and highlights which {DE} models are more robust or least vulnerable to such uncertainties.

The paper is organised as follows. Section \ref{sec:theo} presents the theoretical framework underlying {DE} models. Section \ref{sec:data} describes the observational datasets employed, including SN-Ia and complementary BAO and CMB measurements. Section \ref{sec:method} details the methodology used to infer {DE} parameters. In Section \ref{sec:systematics}, we categorise the systematic effects studied, including calibration, intergalactic dust, progenitor evolution in luminosity and light-curve stretch, increased scatter in color corrections, and density parameter mismatches. Section \ref{sec:results} presents the resulting impact of these systematics on the inferred {DE} constraints. Finally, Section \ref{sec:summary} summarises the main findings and discusses their implications for future surveys and the robustness of {DE} measurements.

\section{Theoretical Framework}
\label{sec:theo}

We consider a spatially flat, homogeneous and isotropic Friedmann–Lema\^itre–Robertson–Walker (FLRW) universe. The expansion of the universe is described by the scale factor $a(t)$, with the Hubble parameter defined as
\begin{equation}
H(t) = \frac{\dot{a}(t)}{a(t)} \, .
\end{equation}

The comoving distance $D_M$ to a source at redshift $z$ in the FLRW universe is
\begin{equation}
D_M(z) = \frac{c}{H_0} \int_{0}^{z} \frac{{\rm d}z'}{E(z')} \, ,
\end{equation}
where $E(z) \equiv H(z)/H_0$ is the normalised Hubble parameter, and the redshift is related to the scale factor by $1+z = a_0/a$. Here, the subscript $0$ indicates the present-day value of the corresponding quantity.

The expansion rate of the universe, described by $H(z)$, depends on the total energy density $\rho_{\rm t}(z)$ through Einstein's equations. This energy density is distributed among different components of the universe, primarily pressureless matter (baryons and cold dark matter) and {DE} with {EoS} $w_{\rm DE}(z) = p_{\rm DE}(z)/\rho_{\rm DE}(z)$. Consequently, $E(z)$ can be written as
\begin{align}
\begin{split}
E^2(z) =~ &\Omega_{m0} (1+z)^3 +  \\
&\left(1 - \Omega_{m0} \right) 
\exp \left[ 3 \int_0^z \frac{1 + w(z')}{1+z'} \, {\rm d}z' \right].
\end{split}
\end{align}

In this work, we consider a set of cosmological models to explore a broad range of systematic uncertainties in {SN-Ia} cosmology, and their impact on the present-day {EoS} $w_0$ and its evolution $w_a$. We test four representative cases:
\begin{enumerate}[left=0pt]
\item {Generalised Scale Factor (GEN) parametrization} \cite{Sen:2001xu, Mukhopadhyay2024}:  
Within the $\Lambda$CDM framework (neglecting radiation at late times), the scale factor has an exact solution,
\begin{equation}
a(t) = a_1 \left[ \sinh\left( {t}/{\tau} \right) \right]^{2/3} \, ,
\end{equation}
where $a_1$ is an arbitrary dimensionless parameter and $\tau$ is a characteristic time scale, yielding
\begin{equation}
E^2(z) = \Omega_{m0} \,  (1+z)^3 + 1 - \Omega_{m0} \, .
\end{equation}
We consider a generalised extension of $a(t)$ to represent new physics, irrespective of any specific {DE} model,
\begin{equation}
a(t) = \tilde{a}_1 \left[ \sinh\left( {t}/{\tau} \right)  \right]^B \, ,
\end{equation}
which leads to
\begin{equation} \label{eq:gen_Ez}
E^2(z) = A\, (1+z)^{2/B} + (1-A) \, , 
\end{equation}
with $B$ a free parameter and $A = \frac{\tilde{a}_1^{2/B}}{\left(1 + \tilde{a}_1^{2/B}\right)}$. The effective {DE} EoS is then
\begin{equation} \label{eq:gen_wz}
w(z) = \frac{w_T(z) \, E^2(z)}{E^2(z) - \Omega_{m0} \, (1+z)^3} \, ,
\end{equation}
where
\begin{equation}
w_T(z) = \frac{2A}{3B} \frac{(1+z)^{2/B}}{E^2(z)} - 1 \, .
\end{equation}
This allows us to directly evaluate $w_0 \equiv w(z=0)$ and $w_a \equiv \left. \frac{{\rm d}w}{{\rm d}z} \right|_{z=0}$, where $w_0$ and $w_a$ denote the present-day EoS and its linear variation, respectively. 

\item {Chevallier–Polarski–Linder (CPL) parametrization} \cite{Chevallier:2000qy, Linder:2002et}:  
The {DE} EoS is modelled as a first-order Taylor expansion in the scale factor $a$,  
\begin{equation}
w(a) = w_0 + w_a \, (1-a) \, ,
\end{equation}
which in terms of redshift $z$ becomes
\begin{equation}
w(z) = w_0 + w_a \, \frac{z}{1+z} \, .
\end{equation}

\item {Jassal–Bagla–Padmanabhan (JBP) parametrization} \cite{Jassal2005}:  
The {DE} evolution peaks at intermediate redshifts and asymptotes to $w_{0}$ at both early and late times,  
\begin{equation}
w(a) = w_{0} + w_{a} \, (1-a) \, a \, ,
\end{equation}
which translates to
\begin{equation}
w(z) = w_0 + w_a \left( \frac{z}{1+z} \right)^2 \, .
\end{equation}

\item {Logarithmic (LOG) parametrization} \cite{Efstathiou1999}:  
This model describes a slowly varying DE EoS, expressed as
\begin{equation}
w(a) = w_{0} - w_{a} \ln a \, ,
\end{equation}
or equivalently,
\begin{equation}
w(z) = w_{0} + w_{a} \, \ln(1+z) \, .
\end{equation}
\end{enumerate}
Together, these four parametrizations allow us to test the robustness of cosmological constraints under different functional forms of $w(a)$.

\section{Observational Data \label{sec:data}}

We utilise the redshift distribution and noise covariance matrix from three key datasets: 
\begin{itemize}[left=0pt]
    \item Pantheon-Plus SN-Ia: The Pantheon-Plus sample consists of $\sim 1500$ spectroscopically-classified SN-Ia, spanning the redshift range $0.01 < z < 2.26$ \cite{Brout2022}. It provides redshift-dependent measurements of the distance modulus $\mu(z)$, 
    \begin{align} \label{eq:mu_cosmo}
    \begin{split}
        \mu(z) &= 5 \log_{10} \left[\frac{D_L(z)}{1\,\mathrm{Mpc}} \right] + 25\, , \\ 
        D_L(z) &= D_M \, (1+z) \, ,    \end{split}
    \end{align} which connects the luminosity distance $D_L(z)$ of a source to its observed flux after accounting for the effects of cosmic expansion and redshift.

    \item DESI-DR2 BAO:  We use 13 correlated {BAO} distance measurements of $D_M/r_d$, $D_H/r_d$, and $D_V/r_d$ from the DESI public data release~2, spanning the redshift range $0.1 < z < 4.2$ (see Table~IV of \cite{DESI:2025zgx}). Here
    \begin{equation}
    D_H(z) = \frac{c}{H(z)} \, ,
    \end{equation}
    is the Hubble distance, and
    \begin{equation}
    D_V(z) = \left[ z \, D_M^2(z) \, D_H(z) \right]^{1/3} \, ,
    \end{equation}
    is the spherically averaged (isotropic) BAO distance. These are expressed in units of the comoving sound horizon at the drag epoch $r_d$. Assuming standard pre-recombination physics, $r_d$ can be approximated as \cite{Brieden:2022heh}  
    \begin{align} 
    r_d & \simeq 147.05 ~\mathrm{ Mpc } ~~ \times \nonumber \\
    &\left( \frac{\omega_b}{0.02236} \right)^{-0.13} 
    \left( \frac{{\omega_{bc}}}{0.1432} \right)^{-0.23} 
    \left( \frac{N_{\rm eff}}{3.04} \right)^{-0.1} \, ,
    \end{align}
where $\omega_b = \Omega_{b} h^2$, {$\omega_{bc} = \omega_{b} + \omega_{c}$, $\omega_{c}= \Omega_{c}h^2$} and $N_{\rm eff}$ is the effective number of relativistic species. 

    \item Compressed CMB: For the CMB, we adopt the compressed likelihood represented by a correlated Gaussian prior on the parameters $\{\theta_\ast, \omega_b, \omega_{bc}\}$. These are set by early-universe physics and can be constrained independently of late-time effects, since contributions from the Integrated Sachs-Wolfe (ISW) effect and CMB lensing have been marginalised. The acoustic scale $\theta_\ast$ is defined as
    \begin{equation}
    \theta_\ast = \frac{r_s(z_\ast)}{D_A(z_\ast)} ,
    \end{equation}
    where $r_s(z_\ast)$ is the sound horizon at recombination and $D_A(z_\ast)$ the corresponding angular diameter distance, {given by $D_A= D_M/(1+z)$}. This compression is based on the \texttt{CamSpec} likelihood implementation, detailed in \cite{Lemos:2023xhs}.
\end{itemize}

For SN-Ia, we generate mock distance moduli at the redshifts of the Pantheon-Plus sample. 
{
The fiducial distance modulus is computed from the luminosity distance defined in Eq.~\eqref{eq:mu_cosmo},
\begin{equation} 
\mu_{\rm fid}(z_i) = 5\log_{10}\left[\frac{D_L^{\rm fid}(z_i)}{\rm Mpc}\right] + 25 , \label{eq:mu_mock}
\end{equation}
where $D_L^{\rm fid}(z_i)$ is evaluated in the assumed fiducial cosmology. The mock SN-Ia distance-modulus vector is then generated as
\begin{equation}
\boldsymbol{\mu}_{\rm mock}
=
\boldsymbol{\mu}_{\rm fid}
+
\boldsymbol{\delta}_{\rm SN},
\qquad
\boldsymbol{\delta}_{\rm SN}
\sim
\mathcal{N}\left(\mathbf{0},\mathcal{C}_{\rm SN}\right),
\end{equation}
where $\mathcal{C}_{\rm SN}$ is the Pantheon-Plus covariance matrix. This covariance includes both statistical and systematic contributions, and is therefore propagated into the mock distance moduli. Residual SN-Ia systematics are subsequently injected as controlled perturbations to $\boldsymbol{\mu}_{\rm mock}$, allowing us to quantify their effect on the recovered cosmological parameters.}

For BAO, we simulate the distance measurements $D_M/r_d$, $D_H/r_d$, and $D_V/r_d$ at the DESI tracer redshifts, adopting the reported covariance to model their correlations. {The mock BAO vector is generated as
\begin{equation}
\small \mathbf{D}_{\rm BAO}^{\rm mock}
=
\mathbf{D}_{\rm BAO}^{\rm fid}
+
\boldsymbol{\delta}_{\rm BAO},
\qquad
\boldsymbol{\delta}_{\rm BAO}
\sim
\mathcal{N}\left(\mathbf{0},\mathcal{C}_{\rm BAO}\right),
\end{equation}
where $\mathbf{D}_{\rm BAO}^{\rm fid}$ contains the fiducial predictions for $D_M/r_d$, $D_H/r_d$, and $D_V/r_d$, and $\mathcal{C}_{\rm BAO}$ is the reported DESI covariance matrix. This propagates the quoted BAO uncertainties and preserves correlations among the tracer measurements.}

For the CMB, we generate synthetic realisations of the compressed parameters $(\theta_\ast, \omega_b, {\omega_{bc}})$ drawn from a multivariate Gaussian comprising the mean vector and covariance matrix, {such that,
\begin{equation}
\small \mathbf{p}_{\rm CMB}^{\rm mock}
=
\mathbf{p}_{\rm CMB}^{\rm fid}
+
\boldsymbol{\delta}_{\rm CMB},
\qquad
\boldsymbol{\delta}_{\rm CMB}
\sim
\mathcal{N}\left(\mathbf{0},\mathcal{C}_{\rm CMB}\right),
\end{equation}
where $\mathbf{p}_{\rm CMB}^{\rm fid}=(\theta_\ast,\omega_b,{\omega_{bc}})_{\rm fid}$ and $\mathcal{C}_{\rm CMB}$ is the covariance matrix of the compressed CMB parameters. This propagates the uncertainties and correlations among the compressed CMB observables.}

The mock data are constructed assuming a fiducial $\Lambda$CDM cosmology with $\Omega_{m0} = 0.3$, $H_0 = 70~\mathrm{km\,s^{-1}\,Mpc^{-1}}$, $w_0 = -1$, and $w_a = 0$ for the CPL, JBP, and LOG parametrizations. For the GEN parametrization, we adopt $A = 0.3$ and $B = 2/3$.  {Since the fiducial cosmology is known by construction, this mock-data framework allows us to isolate and test the impact of residual systematic effects in SN-Ia data.}

\section{Methodology} \label{sec:method}

The observed distance modulus for a {SN-Ia} is obtained from its peak apparent magnitude ($m_B$), light-curve stretch ($x_1$), and colour ($c$) as \cite{Tripp:1997wt, Tripp:1999tc}
\begin{equation} \label{eq:tripp}
\mu_{\mathrm{obs}} = m_B - (M_B - \alpha x_1 + \beta c) + \Delta M + \Delta B,
\end{equation}
where $M_B$ is the absolute magnitude for a fiducial supernova with $x_1 = c = 0$. The parameters $\alpha$ and $\beta$ describe the width–luminosity and colour–luminosity relations, while $\Delta M$ and $\Delta B$ account for host-galaxy luminosity and distance biases \cite{Popovic2021,Popovic2023}.

{In the mock analysis, Eq.~\eqref{eq:tripp} provides the observational motivation for constructing SN-Ia distance moduli, while the actual mock data used in the likelihood are generated from the fiducial luminosity distance relation described in Eq. \eqref{eq:mu_mock} (see Section~\ref{sec:data}). We do not use the real SN-Ia data directly in this controlled-injection part of the analysis because our aim is not to reanalyse the observed supernova sample, but to isolate how specific systematic effects propagate into the inferred cosmological parameters. This is most cleanly achieved with mock data, for which the underlying fiducial cosmology is known, and different systematic effects can be introduced phenomenologically one at a time with prescribed amplitudes.} Cosmological parameter inference uses the full SN-Ia covariance matrix,
\begin{equation}
\mathcal{C}_{\mathrm{SN}} = \mathcal{C}_{\mathrm{stat}} + \mathcal{C}_{\mathrm{sys}},
\end{equation}
with a Gaussian likelihood,
\begin{equation}
\chi^2_{\mathrm{SN}} = \Delta_{\rm SN}^T  \,  \cdot \, \mathcal{C}_{\mathrm{SN}}^{-1} \, \cdot \, \Delta_{\rm SN} \, ,
\end{equation} where,
\begin{equation}
\Delta_{\rm SN}  =  \mu_{\rm mock}(z) - \mu_{\rm model}(z) \, .
\end{equation}

For BAO, the likelihood is constructed from the measured distance ratios $D_X/r_d$ (with $X \equiv M,~ H,~ V$) and their covariance $\mathcal{C}_{\rm BAO}$:
\begin{equation}
\chi^2_{\rm BAO} = \Delta_{\rm BAO}^T \, \cdot \, \mathcal{C}_{\rm BAO}^{-1} \, \cdot \, \Delta_{\rm BAO} \, ,
\end{equation}
where, \begin{equation}
\Delta_{\rm BAO}  =  \left[\frac{D_{X, \, {\rm mock}}}{r_d} - \frac{D_{X, \, {\rm model}}}{r_d}\right] \, .
\end{equation}

For the compressed CMB parameters $\{\theta_\ast, \omega_b, \omega_{bc}\}$, the Gaussian likelihood is defined using the computed mock mean vector $\boldsymbol{\mu}_{\rm CMB}$ and predicted covariance matrix $\mathcal{C}_{\rm CMB}$:
\begin{equation}
\chi^2_{\rm CMB} = \Delta_{\rm CMB}^T \, \cdot \, \mathcal{C}_{\rm CMB}^{-1} \, \cdot \, \Delta_{\rm CMB} \, ,
\end{equation} where, \begin{equation}
\Delta_{\rm CMB} = \{\theta_\ast, \omega_b, {\omega_{bc}}\}_{\rm mock} - \{\theta_\ast, \omega_b, {\omega_{bc}}\}_{\rm model} \, .
\end{equation}

Assuming the datasets are independent, the combined likelihood is the sum of the individual $\chi^2$ contributions:
\begin{equation}
\chi^2_{\rm tot} = \chi^2_{\rm SN} + \chi^2_{\rm BAO} + \chi^2_{\rm CMB}.
\end{equation}
This unified formalism allows simultaneous parameter inference from supernovae, in combination with BAO and CMB observables, while propagating correlations within each dataset.

We perform parameter sampling using the \texttt{emcee}\footnote{\url{https://github.com/dfm/emcee.git}} \cite{emcee}  MCMC ensemble sampler, evaluating the likelihood at each step within a Bayesian framework. Uniform (flat) priors are adopted on all model parameters, with ranges specified in Table~\ref{tab:priors}. The resulting posterior chains are analysed with \texttt{GetDist}\footnote{\url{https://github.com/cmbant/getdist.git}} \cite{Lewis:2019xzd}, which provides marginalized parameter constraints, contour plots, and statistical summaries.

\begin{table}[h]
\centering
\caption{Uniform (flat) priors adopted for cosmological parameters in our analysis. The dimensionless Hubble constant is defined as $h = H_0 / (100~\mathrm{km\,s^{-1}\,Mpc^{-1}})$.}
\label{tab:priors}
\begin{tabular}{lc}
\hline
Parameter ~~~~~~~~~~~~&~~~~~~~~~~~ Prior range \\
\hline
$h$         & $\mathcal{U}\,[0.3, ~ 1.2]$ \\
$\Omega_{m0}$ & $\mathcal{U}\,[0, ~ 1]$ \\
$w_0$       & $\mathcal{U}\,[-3, ~ 1]$ \\
$w_a$       & $\mathcal{U}\,[-3, ~ 2]$ \\
$A$         & $\mathcal{U}\,[0, ~ 1]$ \\
$B$         & $\mathcal{U}\,[0.4, ~ 0.8]$ \\
\hline
\end{tabular}

\vspace{2mm}
\small \textit{Note.} All priors are uniform (flat) within the specified ranges. The chosen intervals are sufficiently broad to avoid constraining the posterior artificially while excluding unphysical values.
\end{table}

\begin{figure*}
    \centering
    \includegraphics[width=\linewidth]{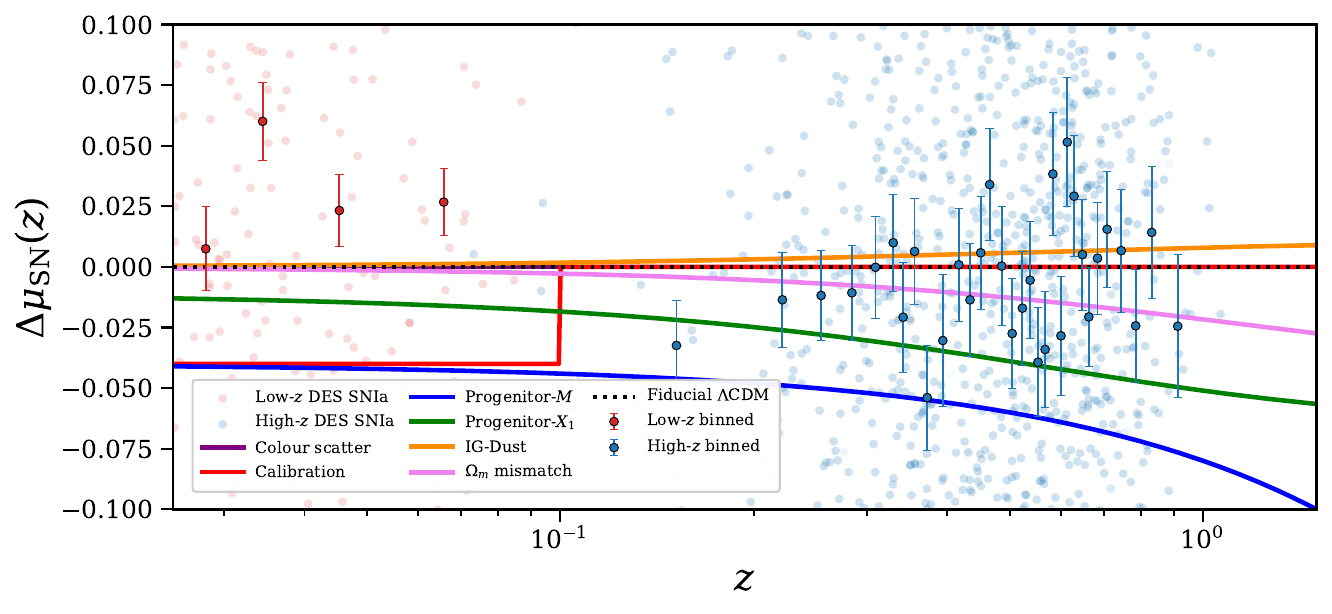}
    \caption{Impact of injected SN-Ia systematic uncertainties on the Hubble residuals, expressed as distance-modulus residuals $\Delta\mu_{\rm SN}(z)$, illustrating how each source of systematics alters the inferred cosmological distances. For comparison, Low-$z$ and High-$z$ DES SN-Ia data are shown, together with their binned mean residuals.}
    \label{fig:injection}
\end{figure*}

\section{Systematic Groupings}
\label{sec:systematics}

Supernova cosmology is sensitive to a variety of systematic effects that can bias the inferred luminosity distances and, consequently, the constraints on {DE}. We classify these effects into broad categories, each motivated by astrophysical or observational considerations. In this section, we outline the physical origin of each source of systematics and the parametrizations we adopt to model their possible redshift dependence.

{\noindent We illustrate in Fig.~\ref{fig:injection} the redshift-dependent Hubble residual shifts produced by the different SN-Ia systematic injections considered in this work, assuming the CPL model. These residuals are defined relative to the fiducial mock SN-Ia distance moduli and show how each systematic modifies the inferred SN-Ia distances as a function of redshift. Any purely redshift-independent component is largely absorbed by the floating SN-Ia normalization associated with $H_0/M_B$; therefore, the relevant effect for DE inference is the redshift-dependent part of the injected residual.}

\subsection{Calibration}
A persistent challenge in SN-Ia cosmology arises from photometric calibration. Because low- and high-redshift SN-Ia are generally observed with different instruments, cross-calibration is required to bring them onto a common flux scale \cite{Brout2022, Vincenzi2024}. The low-redshift anchor sample itself is heterogeneous, having been assembled from multiple telescopes and filter systems, many of which are no longer operational. Recent surveys such as Foundation \cite{Foley_2017}, Zwicky Transient Facility (ZTF) \cite{Dhawan_2021}, and Joint Light-curve Analysis(JLA) + Hubble Space Telescope (HST)/Vera C. Rubin Observatory Legacy Survey of Space and Time (LSST)-pathfinders \cite{Rigault2024}, {and the Dark Energy Bedrock All-Sky Supernova Program (DEBASS) \cite{Sherman:2025rgl,Acevedo:2025fpo}} have sought to provide more uniform low-redshift datasets. {DEBASS is particularly relevant to the present calibration discussion because it uses the same photometric instrument as both DES and the DECam Local Volume Exploration Survey, thereby helping to reduce relative calibration offsets between low-$z$ and high-$z$ samples.} Nonetheless, residual offsets between low- and high-redshift subsets remain possible. 

To capture this effect, we model the calibration systematic as a step function in the $B$-band absolute magnitude:
\begin{equation}
M_B(z) =
\begin{cases} ~
M_B(0) \, , & ~~ z \leq 0.1 \\
~ M_B(0) + \Delta M_B \, , & ~~ z > 0.1
\end{cases}
\end{equation}
where $\Delta M_B$ represents the magnitude offset between low- and high-redshift SN-Ia. 

{As the average difference between low-$z$ and high-$z$ SN-Ia distance moduli is of order $\sim 0.04$ mag \cite{Brout:2021mpj}, we test effects that have a comparable amplitude in the context of $\Lambda$CDM vs $w_0w_a$CDM, as motivated by recent studies \cite{Bansal:2026axl}.}
{In our analysis, we adopt $\Delta M_B=0.02~{\rm mag}$ as a representative residual calibration offset between low- and high-redshift SN-Ia samples. This scale is motivated by the level at which cross-survey photometric calibration and low-redshift anchor calibration remain relevant for current SN-Ia cosmology analyses, even after modern calibration improvements \cite{Scolnic:2013efb, Brownsberger:2021uue, Brout:2021mpj, Acevedo:2025fpo}. The value is therefore not intended to represent a unique calibration error, but rather a controlled test of how a small residual magnitude offset propagates into the inferred DE parameters.} The resulting parameter shifts in the $w_0$–$w_a$ plane for all models are given in Table \ref{tab:result}.

\subsection{Intergalactic Dust}
While host-galaxy dust is routinely corrected for in SN-Ia light-curve analyses, either through empirical colour terms or extinction in the $V$ band, absorption by intergalactic dust is typically neglected. {However, diffuse dust associated with galaxy halos and large-scale structure has been inferred from quasar reddening and magnification measurements, extending to scales of several Mpc from galaxies \cite{Menard2010,Menard2010SN}.}

Current constraints suggest that such a component represents only a tiny fraction of the total baryonic mass \cite{Goobar2018}. Nevertheless, even a low-density diffuse dust population could attenuate SN fluxes in a redshift-dependent manner, thereby biasing the distance-redshift relation and potentially mimicking cosmological effects {\cite{Corasaniti2006,Menard2010SN,Goobar2018}}. In particular, intergalactic dust could shift the inferred {DE} parameters.

Following {previous phenomenological treatments of intergalactic dust attenuation in SN-Ia cosmology \cite{Corasaniti2006,Menard2010SN,Goobar2018}}, we describe the intergalactic dust density as a power-law function of redshift,
\begin{equation}
\rho_\mathrm{dust}(z) = \rho_{\mathrm{dust},0} \, (1+z)^{\gamma} \, ,
\end{equation}
and scale it relative to the critical density $\rho_{c,0} = \frac{3H_{0}^{2}}{8\pi G}$ as
\begin{equation}
\rho_\mathrm{dust}(z)= \rho_{c,0}\,\Omega_\mathrm{dust}(1+z)^{\gamma} \, .
\end{equation}
For our fiducial test, {following \cite{Dhawan:2024gqy}}, we adopt $\Omega_\mathrm{dust} = 5 \times 10^{-6}$ and $\gamma = -1$, corresponding to a very low-density dust component that decreases slowly with redshift. {This normalization is motivated by the galaxy-dust correlation measurement of \cite{Menard2010SN}, which implies a cosmic dust density of $\Omega_\mathrm{dust}\simeq 5\times10^{-6}$ and can bias SN-Ia cosmological constraints at percent level \cite{Menard2010}.}

Assuming the dust is spatially homogeneous but evolves with cosmic time—consistent with detections of diffuse dust several Mpc from galaxies \cite{Menard2010}—the optical depth along the line of sight to a source at redshift $z_s$ is given by
\begin{equation} 
\tau_\lambda = \int_{0}^{z_s} \kappa_\lambda \, \rho_\mathrm{dust}(z) \,
c\,\frac{{\rm d}t}{{\rm d}z}\,{\rm d}z \, .
\end{equation}
Substituting our parametrization yields,
\begin{equation} \small
\tau_\lambda = \rho_{c,0} \, \Omega_\mathrm{dust}
\int_{0}^{z_s} \kappa_\lambda \, \frac{(1+z_s)}{(1+z)} \,
R_{V}(1+z)^{\gamma-1} \, \frac{{\rm d}z}{E(z)} \, .
\end{equation}
Here, $\kappa_\lambda$ denotes the wavelength-dependent mass absorption coefficient, for which we adopt $\kappa_{V} \simeq 1.5 \times 10^{4} \, \mathrm{cm}^{2} \, \mathrm{g}^{-1}$ \cite{WeingartnerDraine2001}. The corresponding shift in observed supernova magnitudes (Hubble residuals) is then given by
\begin{equation}
\Delta_\mathrm{HR} = -2.5 \, \log_{10}\bigl(e^{-\tau_\lambda}\bigr)
\simeq 1.086 \, \tau_\lambda \, .
\end{equation}
The induced shifts in the $w_0$–$w_a$ parameter space due to intergalactic dust are summarised in Table~\ref{tab:result}.

\subsection{Progenitor Evolution in Luminosity}
Spectroscopic studies of  {SN-Ia} at intermediate redshifts ($z \sim 0.5$) and at higher redshifts ($z > 1$) show that their optical properties remain largely consistent across cosmic time \cite{Balland2009, Balland_2018, Dhawan2024Encore}, with only modest deviations detected in the ultraviolet \cite{Ellis2008, Maguire2012, Foley2012UVMismatch} limit. This generic similarity suggests that strong evolutionary effects in SN-Ia spectra are unlikely. However, a mild redshift-dependent drift in their intrinsic luminosity cannot be fully ruled out{.} {Note that progenitor evolution and intergalactic dust can both introduce redshift-dependent trends in the SN-Ia Hubble diagram, and may therefore lead to partially degenerate phenomenological signatures in the inferred cosmological parameters. Their physical origins, however, are distinct. Intergalactic dust attenuates the observed flux along the line of sight, whereas progenitor evolution changes the intrinsic SN-Ia luminosity or light-curve properties entering the standardization relation. In this work, we therefore treat them as separate systematic templates to isolate their individual impact on the recovered DE constraints.}

To test for such an effect, we introduce a simple phenomenological model in which the absolute $B$-band magnitude evolves with redshift, as
\begin{equation}
M_B(z) = M_B(0) + \epsilon \,(1+z_{\mathrm{CMB}}),
\end{equation}
where $\epsilon$ quantifies the slope of the luminosity evolution.

{
For our analysis, we adopt $\epsilon = 0.02$~mag as a representative phenomenological amplitude for a mild residual luminosity drift. This value is not intended as a unique physical prediction, but as a controlled test of small redshift-dependent evolution in the SN-Ia luminosity \cite{Nordin:2010ti,Dhawan2024Encore}.} This framework allows us to assess whether gradual changes in progenitor properties, such as metallicity or stellar age, producing a mild redshift-dependent shift in intrinsic SN-Ia brightness, could lead to systematic biases in the inferred cosmological parameters. See Table~\ref{tab:result} for the resulting parameter shifts.

\subsection{Progenitor Evolution in Light-Curve Stretch}
The observable properties of {SN-Ia} depend on progenitor characteristics, such as metallicity \cite{MorenoRaya2016a,MorenoRaya2016b} and stellar population age \cite{Childress2014}, which evolve with redshift. These changes can influence both intrinsic luminosity and light-curve shapes, particularly the stretch parameter $x_1$. Recent studies indicate that the width–luminosity correction may differ between low-$x_1$ and high-$x_1$ populations \cite{Ginolin_2025}, implying that a redshift-dependent $x_1$ distribution could bias cosmological inferences if neglected.

We model this effect using a broken power-law for the stretch–luminosity coefficient $\alpha$: \[
\alpha(x_1) =
\begin{cases}
\alpha_\mathrm{low~} = 0.23\, , \hfill & x_1 < -0.49\, , \\
\alpha_\mathrm{high} = 0.13\, , \hfill & x_1 \ge -0.49\, .
\end{cases}
\]
This approach approximates the situation in which selection effects have been corrected, and the remaining offset between true and inferred distances arises from applying a single $x_{1}$-independent $\alpha$ to a population whose $x_{1}$ distribution evolves with redshift. {The numerical values of $\alpha_{\rm low}$ and $\alpha_{\rm high}$ are chosen to follow the two-population stretch-luminosity relation reported by \cite{Ginolin_2025}, and are used here as a controlled test of the bias that would arise if such population dependence were absorbed into a single effective stretch correction.}

To model the redshift evolution of $x_{1}$, we follow the prescription of \cite{Nicolas2021}, which links the shift in $x_{1}$ to the local specific star-formation rate. We write the Hubble residual shift as
\begin{equation}
\Delta_\mathrm{HR} = \alpha(x_{1})\,\Delta x_{1}\, ,
\end{equation}
where $\alpha(x_{1})$ is the stretch–luminosity slope and $\Delta x_{1}$ describes the redshift evolution of the stretch parameter. We parameterize $\Delta x_{1}$ as
\begin{equation}
\Delta x_{1} = F\,\mu_{1} + \left( 1-F\right) \left[a\,\mu_{1} + (1-a)\,\mu_{2}\right] \, ,
\end{equation}
with $a=0.47$, $\mu_{1}=0.38$, and $\mu_{2}=-1.26$, representing the conservative case from \cite{Nicolas2021}. {We adopt this conservative case to avoid overstating the effect of stretch-population evolution while retaining a literature-motivated redshift dependence for the underlying $x_1$ distribution.}

The factor $F$ evolves with redshift as
\begin{equation}
F(z)=\left[K^{-1}(1+z)^{2.8}+1\right]^{-1} \, ,
\end{equation}
which captures the population drift similar to the evolution predicted by \cite{Childress2014}. While alternative parameter choices may modify the detailed evolution, this prescription provides a realistic baseline for testing the influence of progenitor evolution on the $w_{0}$–$w_{a}$ constraints (see Table \ref{tab:result}). This framework allows us to quantify the bias introduced into the cosmological parameters if the evolving $x_{1}$ distribution is not fully accounted for.

\begin{figure*}
    \centering
    \includegraphics[width=\linewidth]{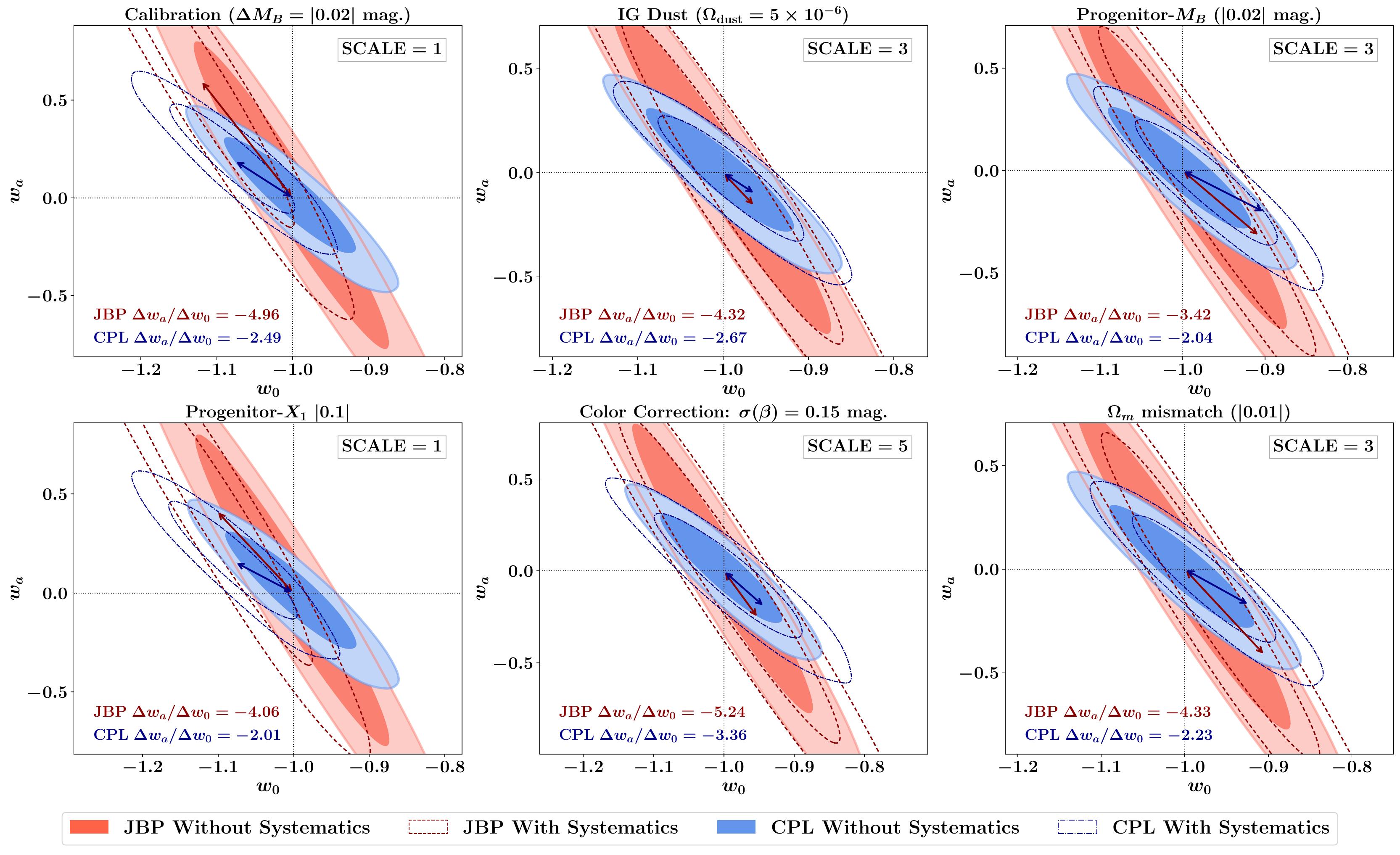}
\caption{{Impact of injected SN-Ia systematics on the $w_0-w_a$ constraints for the CPL and JBP parametrizations. Each panel shows the shift produced by one injected systematic. The panels follow the order in which the systematics are introduced in Section~\ref{sec:systematics}: calibration, intergalactic dust, progenitor evolution in $M_B$, progenitor evolution in $x_1$, colour-correction scatter, and $\Omega_m$ mismatch. The baseline and systematic-injected constraints are distinguished by colour and line style, with some shifts scaled for better visualisation.}}
    \label{fig:jbp_vs_cpl}
\end{figure*}

\begin{figure*}
    \centering
    \includegraphics[width=\linewidth]{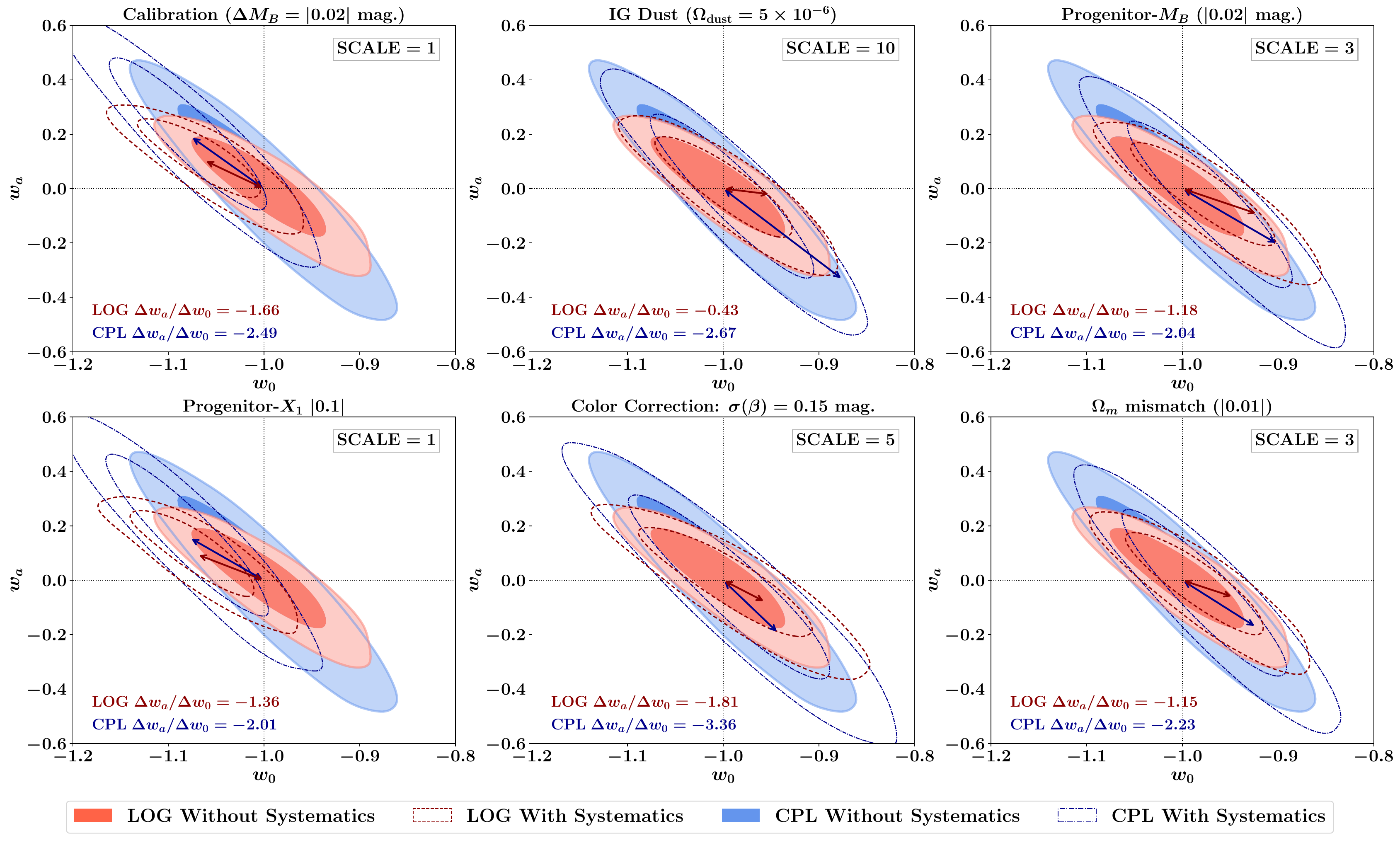}
\caption{{Impact of injected SN-Ia systematics on the $w_0-w_a$ constraints for the CPL vs LOG parametrizations. The panel order and contour conventions are the same as in Fig.~\ref{fig:jbp_vs_cpl}.}}
    \label{fig:log_vs_cpl}
\end{figure*}

\begin{figure*}
    \centering
    \includegraphics[width=\linewidth]{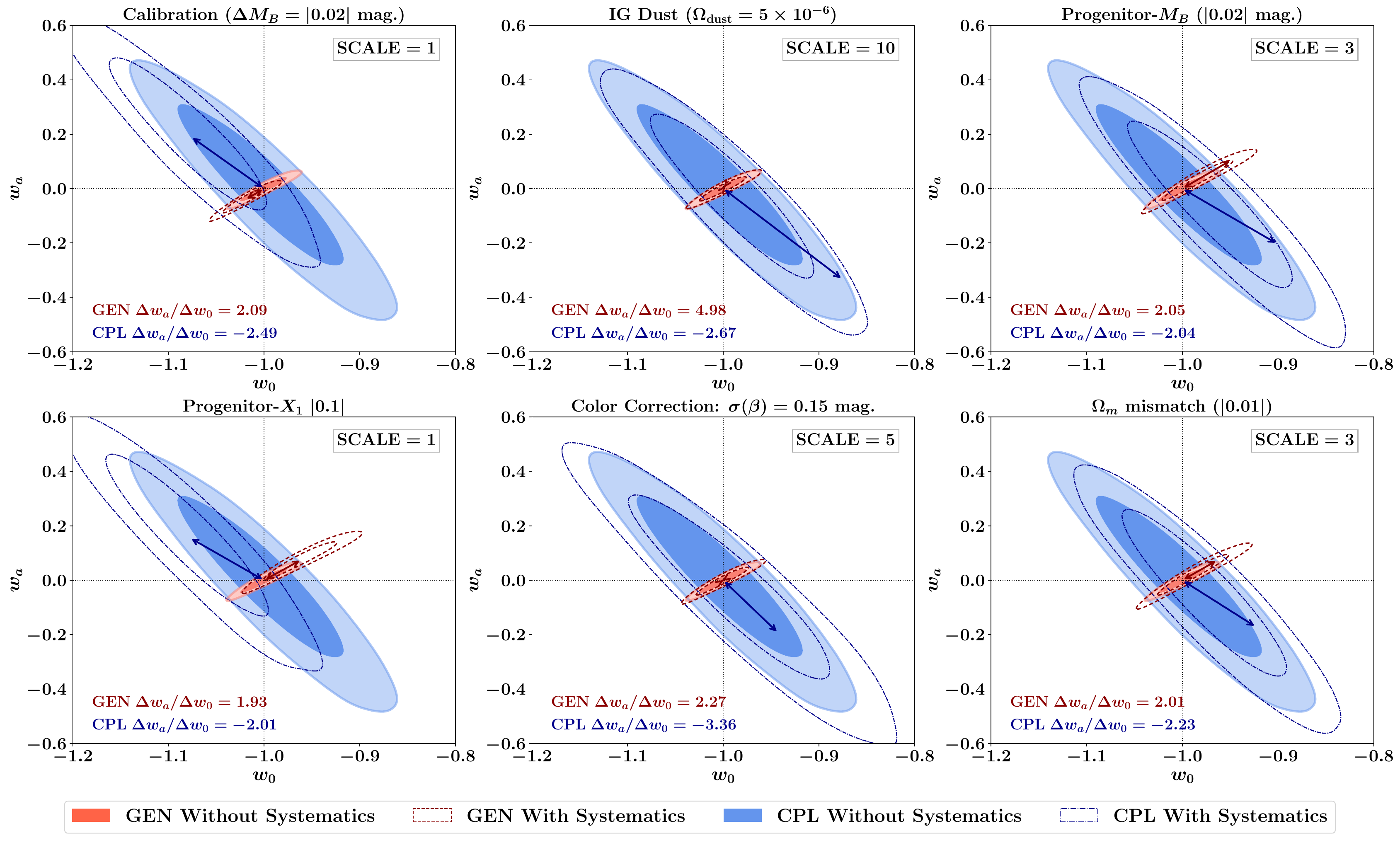}
\caption{{Impact of injected SN-Ia systematics on the $w_0-w_a$ constraints for the CPL vs GEN parametrizations. The panel order and contour conventions are the same as in Fig.~\ref{fig:jbp_vs_cpl}.}}
    \label{fig:gen_vs_cpl}
\end{figure*}

\begin{table}
\centering
\caption{{Absolute shifts in the recovered mean parameters for different SN-Ia systematic injections and DE models, relative to the corresponding no-systematics mock fit.}}
\renewcommand{\arraystretch}{1.2}
\setlength{\tabcolsep}{3.8pt}
\begin{tabular}{lcccc}
\toprule
\multicolumn{5}{c}{\centering \bf Calibration Offset: $\Delta M_B = 0.02$} \\
\hline
\textit{Parameter Shift} & \textbf{CPL} & \textbf{JBP} & \textbf{LOG} & \textbf{GEN} \\
\midrule
$\Delta A$ & --- & --- & --- & $-0.0181$ \\
$\Delta B$ & --- & --- & --- & $-0.0100$ \\
$\Delta w_0$ & $-0.0763$ & $-0.1208$ & $-0.0622$ & $-0.0197$ \\
$\Delta w_a$ & $+0.1905$ & $+0.5988$ & $+0.1034$ & $-0.0411$ \\
\midrule
\multicolumn{5}{c}{\centering \bf Intergalactic Dust Modeling: $\Omega_\mathrm{dust} = 5\times10^{-6}$} \\
\hline
\textit{Parameter Shift} & \textbf{CPL} & \textbf{JBP} & \textbf{LOG} & \textbf{GEN} \\
\midrule
$\Delta A$ & --- & --- & --- & $+0.0002$ \\
$\Delta B$ & --- & --- & --- & $-6.8\times10^{-5}$ \\
$\Delta w_0$ & $+0.0125$ & $+0.0123$ & $+0.0049$ & $+0.0003$ \\
$\Delta w_a$ & $-0.0333$ & $-0.0532$ & $-0.0021$ & $+0.0015$ \\
\midrule
\multicolumn{5}{c}{\centering \bf Progenitor Evolution in $M_B$: $\epsilon = 0.02$} \\
\hline
\textit{Parameter Shift} & \textbf{CPL} & \textbf{JBP} & \textbf{LOG} & \textbf{GEN} \\
\midrule
$\Delta A$ & --- & --- & --- & $+0.0135$ \\
$\Delta B$ & --- & --- & --- & $+0.0064$ \\
$\Delta w_0$ & $+0.0331$ & $+0.0311$ & $+0.0262$ & $+0.0171$ \\
$\Delta w_a$ & $-0.0676$ & $-0.1064$ & $-0.0310$ & $+0.0349$ \\
\midrule
\multicolumn{5}{c}{\centering \bf Progenitor Evolution in $x_1$} \\
\hline
\textit{Parameter Shift} & \textbf{CPL} & \textbf{JBP} & \textbf{LOG} & \textbf{GEN} \\
\midrule
$\Delta A$ & --- & --- & --- & $+0.0348$ \\
$\Delta B$ & --- & --- & --- & $+0.0168$ \\
$\Delta w_0$ & $-0.0777$ & $-0.1019$ & $-0.0698$ & $+0.0391$ \\
$\Delta w_a$ & $+0.1562$ & $+0.4140$ & $+0.0947$ & $+0.0756$ \\
\midrule
\multicolumn{5}{c}{\centering \bf Color Correction: $\sigma_\beta = 0.15$} \\
\hline
\textit{Parameter Shift} & \textbf{CPL} & \textbf{JBP} & \textbf{LOG} & \textbf{GEN} \\
\midrule
$\Delta A$ & --- & --- & --- & $-1.2\times10^{-4}$ \\
$\Delta B$ & --- & --- & --- & $-3.4\times10^{-5}$ \\
$\Delta w_0$ & $+0.0115$ & $+0.0097$ & $+0.0088$ & $-0.0001$ \\
$\Delta w_a$ & $-0.0387$ & $-0.0509$ & $-0.0159$ & $-0.0003$ \\
\midrule
\multicolumn{5}{c}{\centering \bf Density Parameter Mismatch: $\Delta\Omega_{m} = 0.01$} \\
\hline
\textit{Parameter Shift} & \textbf{CPL} & \textbf{JBP} & \textbf{LOG} & \textbf{GEN} \\
\midrule
$\Delta A$ & --- & --- & --- & --- \\
$\Delta B$ & --- & --- & --- & --- \\
$\Delta w_0$ & $+0.0257$ & $+0.0319$ & $+0.0177$ & $+0.0123$ \\
$\Delta w_a$ & $-0.0572$ & $-0.1381$ & $-0.0203$ & $+0.0247$ \\
\bottomrule
\end{tabular}
\label{tab:result}
\end{table}

\begin{table*}
\centering
\caption{{Covariance-weighted tension $T_\sigma$ from the fiducial $(w_0=-1,\, w_a=0)$ for mock-injection cases.}}
\renewcommand{\arraystretch}{1.25}
\setlength{\tabcolsep}{9pt}
\begin{tabular}{lcccc}
\toprule
\textbf{Systematic case} & \textbf{CPL} & \textbf{JBP} & \textbf{LOG} & \textbf{GEN} \\
\midrule
Calibration offset, $\Delta M_B=0.02~{\rm mag}$ & 1.53 & 1.70 & 1.53 & 1.36 \\
Intergalactic dust, $\Omega_{\rm dust}=5\times10^{-6}$ & 0.24 & 0.20 & 0.19 & 0.10 \\
Progenitor evolution in $M_B$, $\epsilon=0.02~{\rm mag}$ & 0.76 & 0.65 & 0.72 & 0.72 \\
Progenitor evolution in $x_1$ & 1.84 & 1.80 & 1.91 & 1.63 \\
Colour-correction scatter, $\sigma_\beta=0.15~{\rm mag}$ & 0.17 & 0.09 & 0.15 & 0.01 \\
Density-parameter mismatch, $\Delta\Omega_m=0.01$ & 0.55 & 0.51 & 0.51 & 0.50 \\
\bottomrule
\end{tabular}
\label{tab:tension_sigma}
\end{table*}

\begin{figure*}
    \centering
    \includegraphics[width=0.7\textwidth]{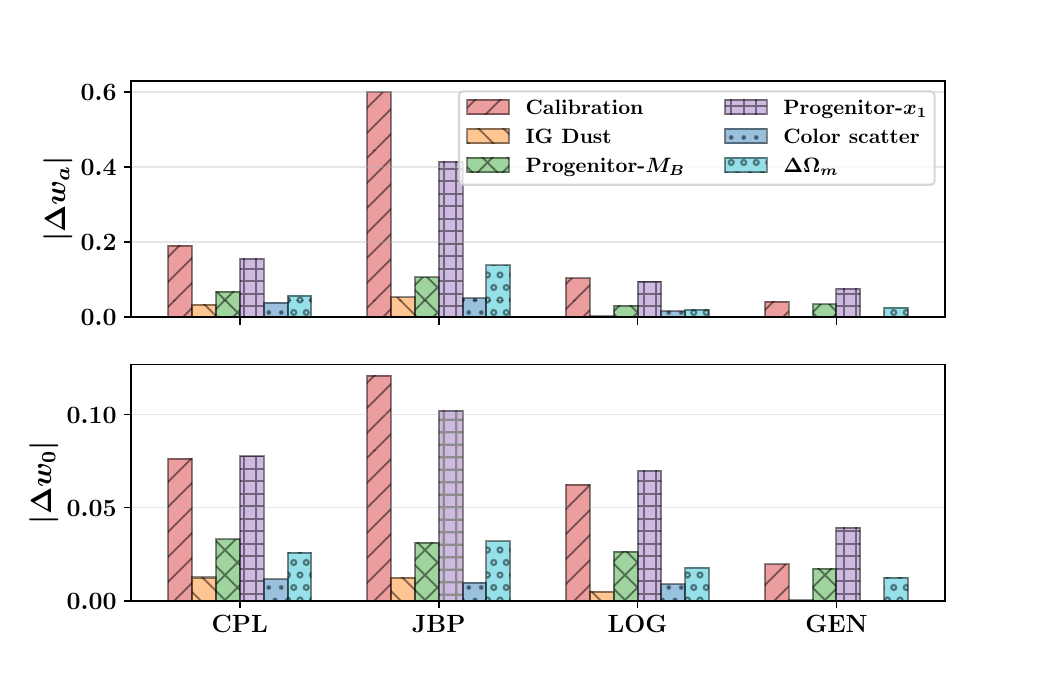}
\caption{{Visual summary of the absolute parameter shifts reported in Table~\ref{tab:result}. The upper panel shows $|\Delta w_a|$ and the lower panel shows $|\Delta w_0|$ for the CPL, JBP, LOG, and GEN parametrizations, with separate bars for the different systematic effects.}}
\label{fig:graph}
\end{figure*}

\subsection{Increased Scatter in Colour Correction}
While systematic offsets primarily shift the relative distance modulus between low- and high-$z$ {SN-Ia}, additional scatter {mainly increases the uncertainties and is not expected to introduce a large coherent mean shift.} We model this effect by introducing extra dispersion in the colour-luminosity correction term ($\beta c$ in the Tripp relation in Eq. \eqref{eq:tripp}), represented as a systematic $\sigma(\beta)$. This captures potential variations in either the dust law slope or the intrinsic colour-luminosity relation. 

In practice, we implement this effect by inflating the covariance matrix. Specifically, we add a diagonal component corresponding to a constant error of $\sigma(\beta) \equiv \sigma_\beta = 0.15$~mag \cite{Foley_2017, Brout2022}, such that  
\begin{equation}
\mathcal{C}' = \mathcal{C} + \sigma_\beta^2 \, \mathrm{I} ,   
\end{equation}
where $\mathcal{C}$ is the original covariance matrix and $\mathrm{I}$ is the identity matrix of dimension equal to the number of supernovae. The updated covariance $\mathcal{C}'$ is then utilised in the likelihood analysis. Although motivated here as colour-correction scatter, this framework can equally represent other forms of residual dispersion in the distance modulus. The impact on the $w_0$–$w_a$ constraints for all models is summarized in Table~\ref{tab:result}.

\subsection{Density Parameter Mismatch} \label{sec:omega_m_mismatch}
Current constraints on the {DE} parameters rely on a joint analysis of early- and late-universe probes, namely the CMB, BAO, and SN-Ia. Within a flat $\Lambda$CDM framework, $(\Omega_{k0}, \, w_0, \,  w_a) = (0, \, -1, \,  0)$, a mild but notable tension arises: the matter density inferred from Planck $\Omega_{m0} = 0.315 \pm 0.007$ \cite{Planck:2018vyg} is lower than that inferred from DES SN-Ia $\Omega_{m0} = 0.357 \pm 0.017$ \cite{DES:2024jxu}, a discrepancy at the $\sim 2.5\sigma$ level. Because $\Omega_{m0}$ is degenerate with $w_0$ and $w_a$ in combined CMB+BAO+SN analyses, such a mismatch can shift the inferred confidence contours compared to the case where both early- and late-time probes favour the same $\Omega_{m0}$. To mimic this effect, we simulate a deliberate offset between the $\Omega_{m0}$ used to generate the SN-Ia Hubble diagram and the value adopted for the external datasets.  A recent study by \cite{Tang:2024lmo} explored how underlying dynamical {DE} cosmologies can induce biases in $\Omega_{m0}$ and $H_0$ when individual probes are analyzed under $\Lambda$CDM. In contrast to their work, our analysis examines the impact of such systematic offsets within the SN-Ia Hubble diagram and evaluates their propagation to the inferred $(w_0, w_a)$ parameters across multiple {DE} parametrizations. 

While the physical origin of such a discrepancy could stem from unresolved systematics in any of the previously discussed categories, here we encapsulate it simply as an $\Omega_{m0}$ mismatch. {This treatment is intended as a proxy for what one can refer to as \textit{unknown unknowns}, i.e., residual or unmodelled sources of bias in the SN-Ia Hubble diagram whose physical origin is not specified in advance. Such effects may arise from combinations of calibration, selection, population-evolution, or environmental systematics that are not captured by any single template considered above. We therefore use the $\Omega_{m0}$ mismatch as a phenomenological way to test how a coherent distortion of the observed $\mu-z$ relation can propagate into the inferred cosmological parameters.}

For the CPL, JBP, and LOG parametrizations, we implement this effect directly as an offset in $\Omega_{m0}$. In contrast, for the GEN parametrization, which is formulated in terms of the parameters $A$ and $B$, the same mismatch is accounted for when transferring from $\{A, B\}$ to $\{w_0, w_a\}$, where $\Omega_{m0}$ enters implicitly through Eqs.~\eqref{eq:gen_Ez} and \eqref{eq:gen_wz}. This allows the resulting $(w_0, w_a)$ shifts to reflect the corresponding density mismatch within the GEN framework.

{Figure \ref{fig:injection} therefore provides a useful visual guide to how the injected systematics enter the Hubble diagram before their impact is propagated into the recovered DE parameters. The calibration offset produces an approximately step-like shift between the low- and high-redshift samples, while progenitor evolution and intergalactic dust introduce smoother redshift-dependent trends. The colour-scatter case primarily increases the effective dispersion rather than producing a coherent mean offset. }

\begin{figure*}
    \centering
    \includegraphics[width = \linewidth]{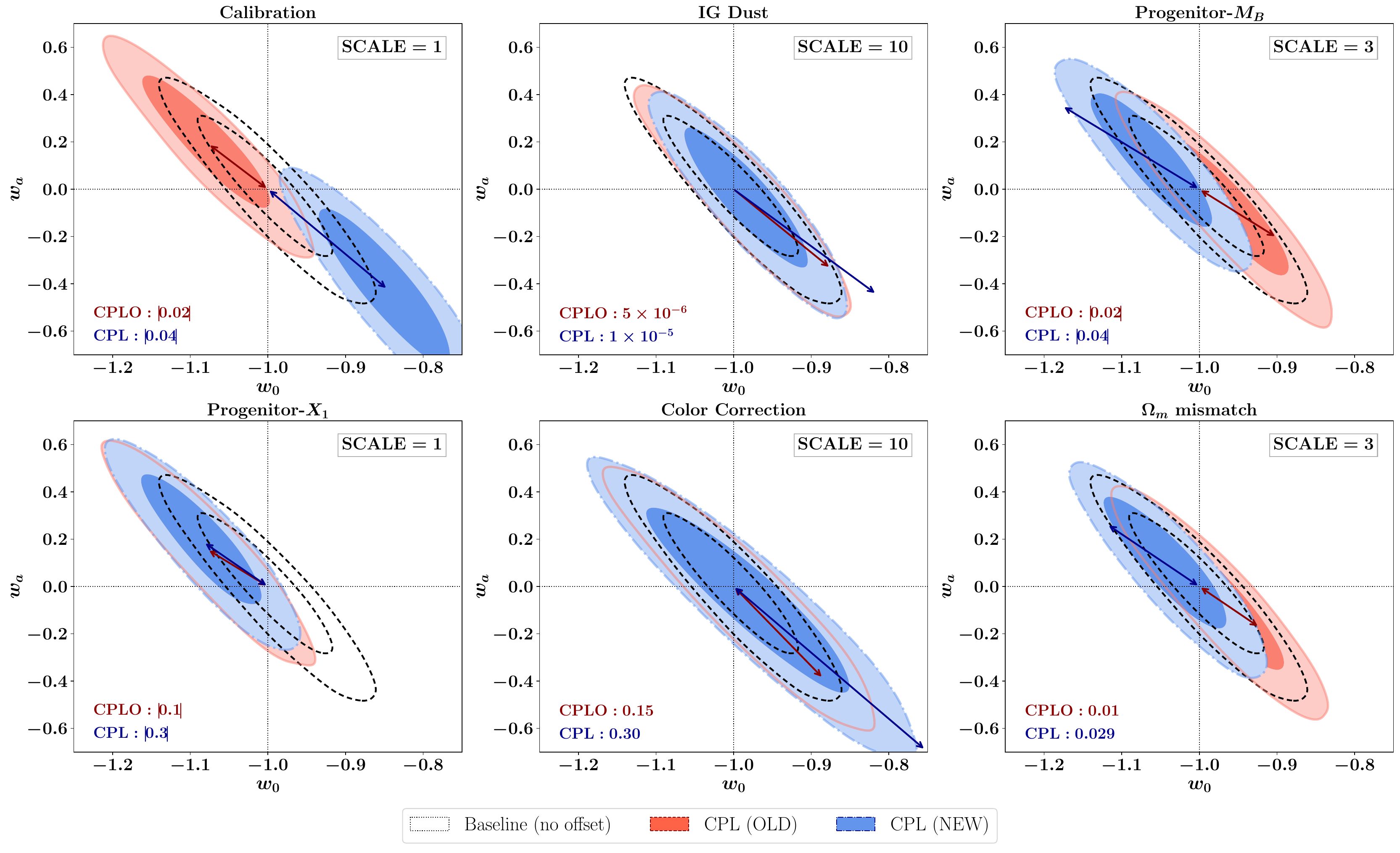}
    \caption{{Impact of injected SN-Ia systematics on the  $w_0-w_a$ constraints for the CPL parametrizations. Panels illustrate how individual systematics shift the recovered contours for two different injected values of each systematic. Dashed contours denote the fits without systematics; filled contours show results after systematic injection.  }}
    \label{fig:cpl_old_vs_new}
\end{figure*}

\begin{figure*}
    \centering
    \includegraphics[width = \linewidth]{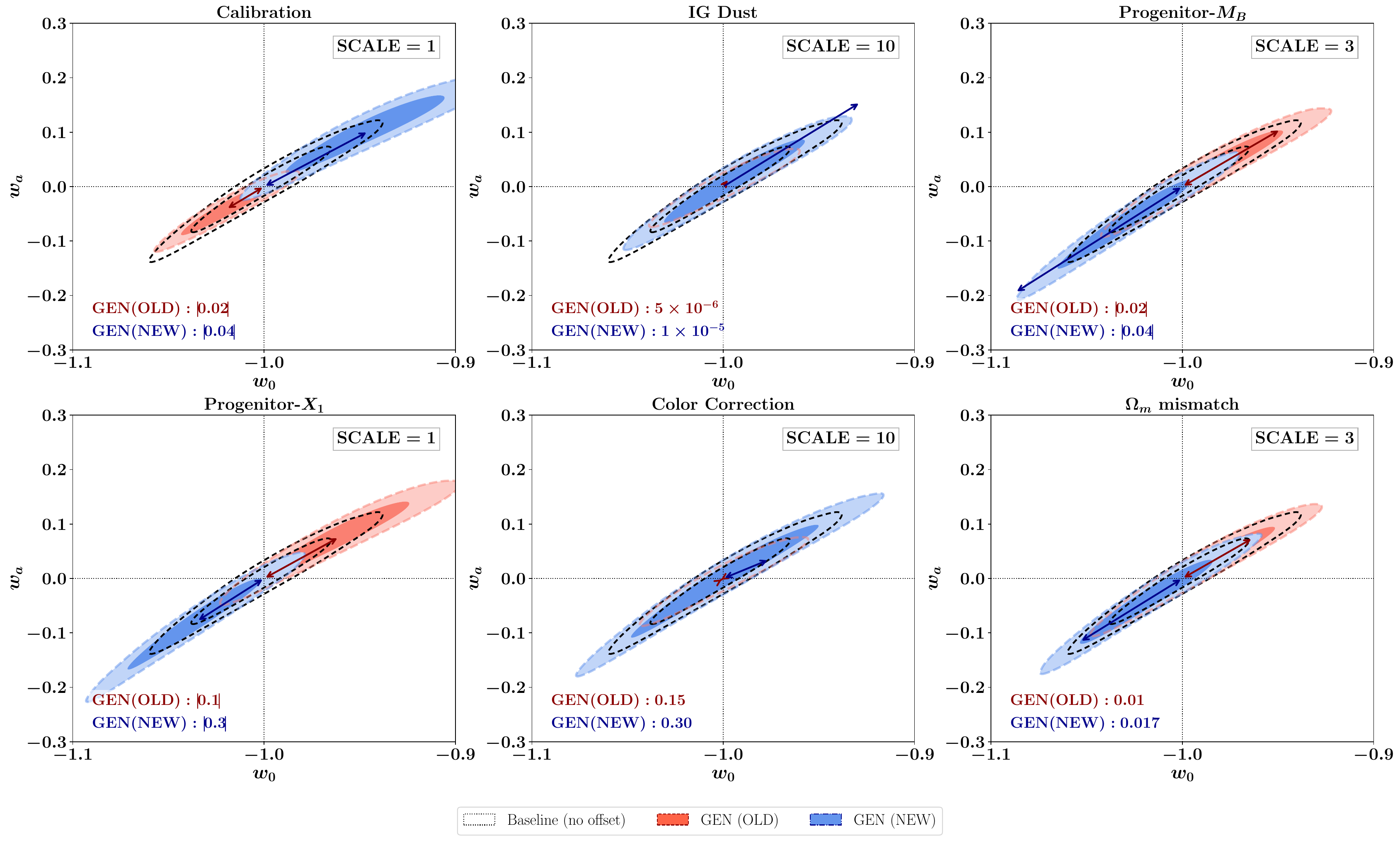}
    \caption{{Impact of injected SN-Ia systematics on the  $w_0-w_a$ constraints for the GEN parametrizations. }}
    \label{fig:gen_old_vs_new}
\end{figure*}

\begin{figure*}
    \centering
    \begin{subfigure}{0.8\linewidth}
        \centering
    \includegraphics[width=0.47\linewidth]{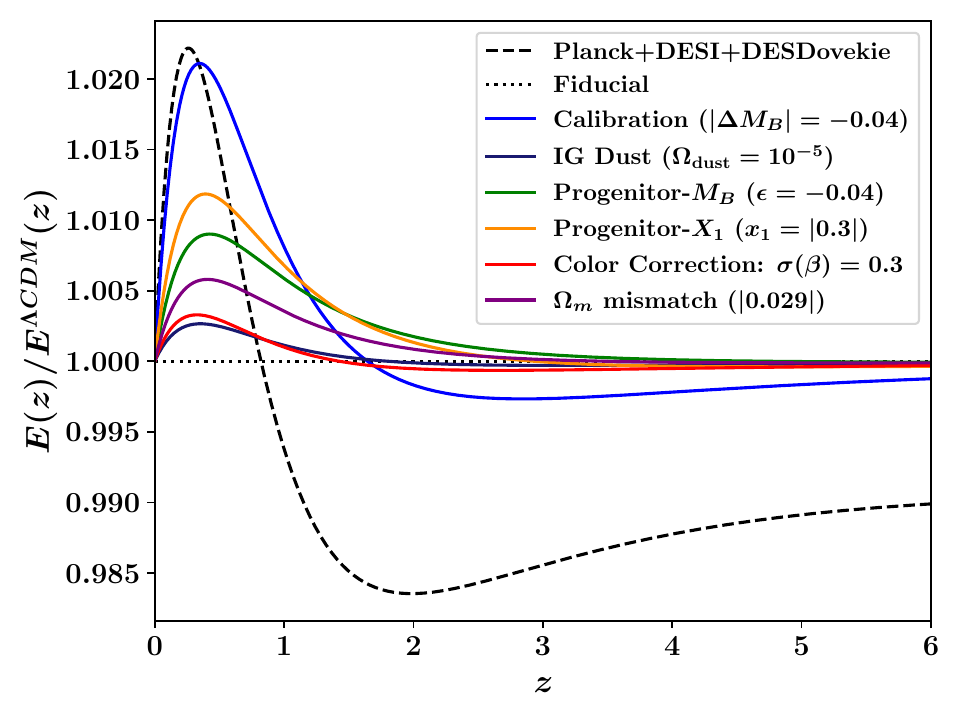}
        \hfill
    \includegraphics[width=0.47\linewidth]{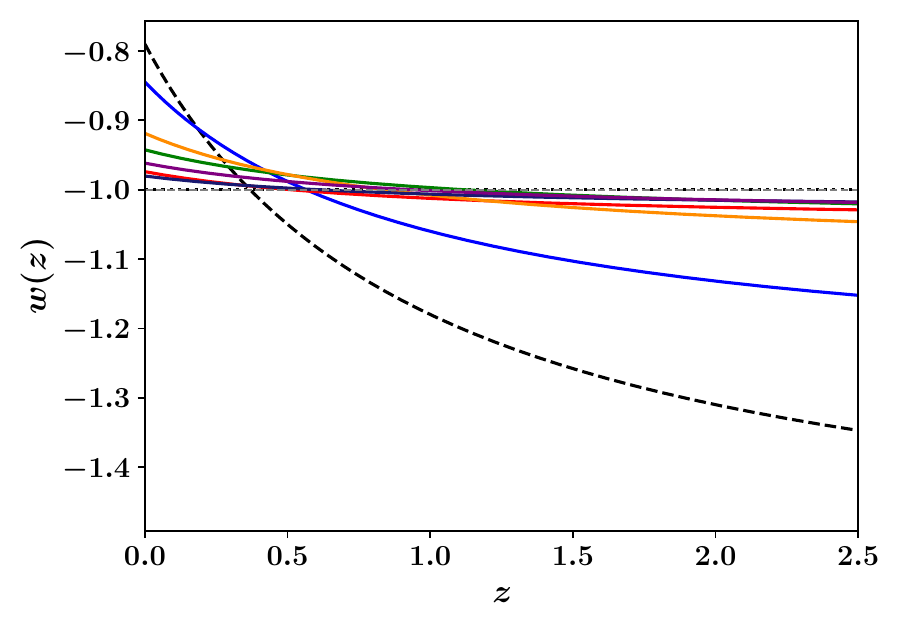}
    \end{subfigure}
    \begin{subfigure}{0.8\linewidth}
        \centering
    \includegraphics[width=0.47\linewidth]{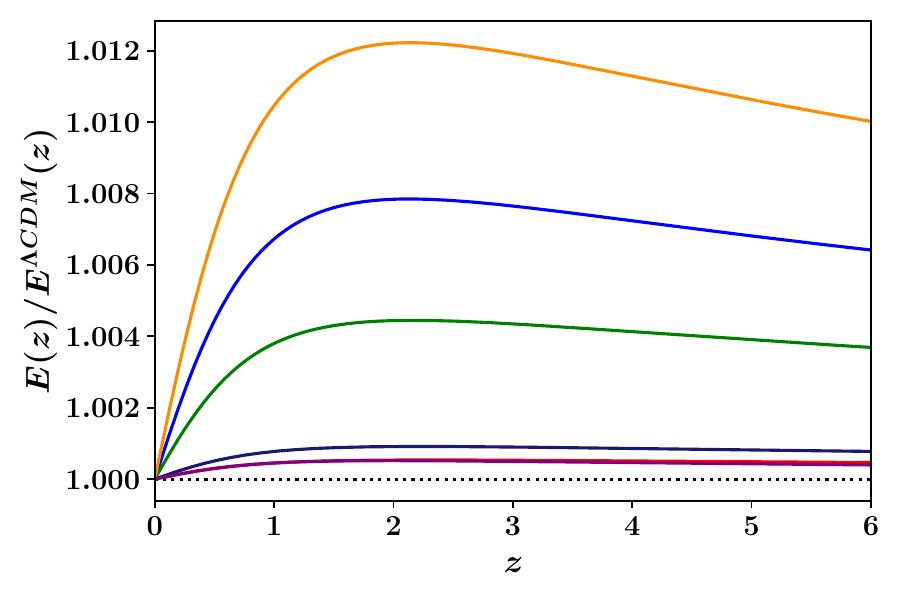}
        \hfill
    \includegraphics[width=0.47\linewidth]{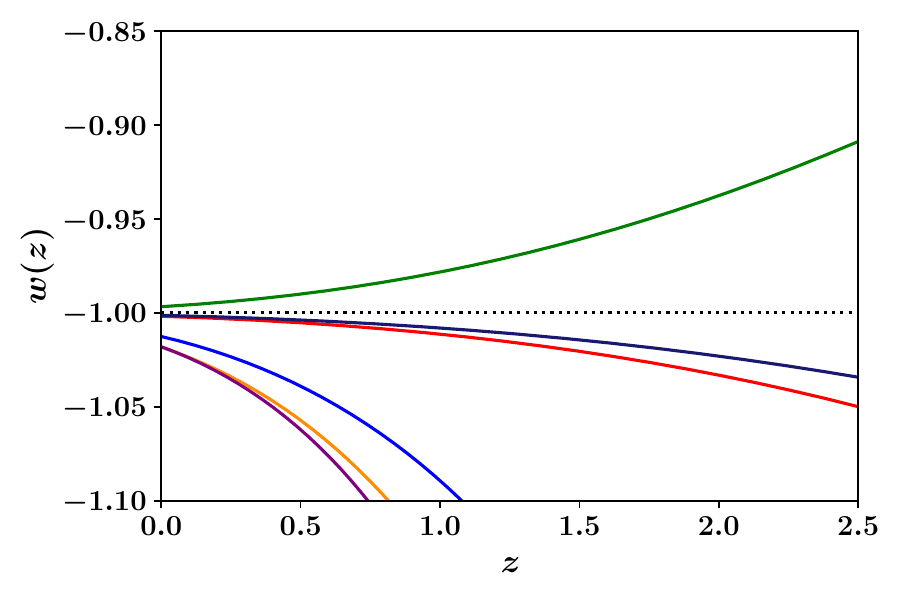}
    \end{subfigure}
\caption{Effect of systematics injection on the normalised Hubble parameter $E(z)$ (left), and {DE} EoS $w(z)$ (right): Comparison between CPL {(top panels)} vs GEN {(bottom panels)} models.}
\label{fig:residual}
\end{figure*}

\section{Impact on Dark Energy}
\label{sec:results}

The influence of the six systematic effects described in Section~\ref{sec:systematics} is quantified in terms of their induced shifts in the inferred {DE EoS} parameters $w_0$ and $w_a$. Table~\ref{tab:result} summarises the parameter displacements for the four {DE} models considered, viz. CPL, JBP, LOG, and GEN, while Figures~\ref{fig:jbp_vs_cpl}, \ref{fig:log_vs_cpl} and \ref{fig:gen_vs_cpl} illustrate their direction and magnitude in the $w_{0}$--$w_{a}$ plane for the four parametrizations studied here.

The shifts reported in Table~\ref{tab:result} denote the absolute displacement of the recovered mean values in the $w_0-w_a$ plane after each systematic injection, measured relative to the corresponding no-systematics mock fit for the same parametrization, generated with fiducial values $w_0=-1$ and $w_a=0$. {This comparison should be interpreted as a common diagnostic in the effective $(w_0,w_a)$ plane, rather than as a direct comparison of identically defined model parameters. The CPL form corresponds to the conventional $w_0w_a$CDM parametrization, while JBP and LOG are alternative two-parameter EoS extensions with different redshift dependences. For GEN, the primary fitted parameters, $(A,B)$, are mapped into effective $(w_0,w_a)$ values through a first-order expansion of the EoS, in analogy with the CPL form.}

{To quantify these displacements while accounting for the $w_0-w_a$ covariance and degeneracy direction, we compute the covariance-weighted tension,
\begin{equation}
T_\sigma =
\left[
\Delta \mathbf{p}^{T}
\mathcal{C}^{-1}
\Delta \mathbf{p}
\right]^{1/2},
\end{equation}
where $\Delta \mathbf{p}=(\Delta{w}_0,\Delta{w}_a)^\text{T}$ and $\mathcal{C}$ is the marginalized covariance matrix of $(w_0,w_a)$ from the MCMC chain. Larger values of $T_\sigma$ indicate a larger deviation from the fiducial model.
}

{To make the comparison in Table~\ref{tab:result} easier to interpret, Fig.~\ref{fig:graph} presents the same information as absolute shifts in the recovered parameters. The upper panel shows $|\Delta w_a|$ and the lower panel shows $|\Delta w_0|$, with the four DE parametrizations placed along the horizontal axis and the different systematic effects shown separately. This visual summary, together with the tension metrics in Table \ref{tab:tension_sigma} highlights that calibration and progenitor-related effects dominate the shifts, while intergalactic dust, colour-correction scatter, and density-parameter mismatch produce smaller displacements.}

\subsection{Effect of Individual Systematics Injection}

The six systematic scenarios lead to distinct patterns of parameter displacement, reflecting their physical origins and the structure of parameter degeneracies. Across most cases, the induced displacements align with the existing $w_{0}$--$w_{a}$ degeneracy direction, indicating that systematics primarily bias the inferred evolution of {DE} rather than its overall normalisation.
\begin{enumerate}[left=0pt]
    \item A uniform magnitude bias of $\Delta M_{B}=0.02$ produces one of the most pronounced effects across all models. The shift is maximum for the JBP parametrization with $\Delta w_{0}\simeq-0.12$ and $\Delta w_{a}\simeq+0.60$, whereas it remains mild for GEN ($\Delta w_{0}\simeq-0.02$, $\Delta w_{a}\simeq-0.04$). This behaviour reflects the strong sensitivity of luminosity distances to global calibration, which directly impacts the absolute brightness scale of the SN-Ia sample.
    \item A mild intrinsic luminosity evolution, parametrized by $\epsilon=0.02$, causes modest but coherent shifts in the $w_0-w_a$ plane. Typical values are $\Delta w_0\sim+0.03$ and $\Delta w_a\sim-0.07$ for CPL, with slightly smaller amplitudes for GEN. Although small in magnitude, this effect introduces a redshift-dependent bias that can imitate {DE} evolution, particularly in more flexible models like JBP.
    \item Evolution in the stretch distribution, linked to host properties and star-formation rate, 
    {for the adopted amplitudes, produces a larger shift than progenitor evolution in $M_B$. This comparison should be understood within the specific normalization choices used here, since the relative impact of different systematics depends on both their amplitude and redshift dependence.} It drives parameter displacements up to $\Delta w_0\sim-0.08$ and $\Delta w_a\sim+0.40$ for the JBP model, again oriented along the degeneracy axis. Among {the progenitor-related templates considered here}, this mechanism contributes the largest bias, highlighting the need to model population drift in SN standardization.
    \item For a fiducial intergalactic dust density of $\Omega_{\mathrm{dust}}=5\times10^{-6}$, the resulting shifts are minimal ($|\Delta w_0|<0.02$, $|\Delta w_a|<0.05$) for all models. {Although the intergalactic-dust model is explicitly redshift dependent, the adopted low normalization produces only a weak, smooth dimming across the redshift range considered here. As a result, its impact on the recovered DE parameters remains small compared with calibration and progenitor-evolution effects.}
    \item Uncertainty in the colour–luminosity coefficient ($\sigma_\beta=0.15$) broadens the error contours but yields negligible mean displacement ($|\Delta w_0|<0.02$, $|\Delta w_a|<0.05$).  Its effect is statistically similar to increased photometric noise, primarily enlarging parameter uncertainties rather than significantly shifting the central estimates.
    \item A shift of $\Delta\Omega_m=0.01$ between the true and assumed cosmology induces small but coherent offsets in $w_0$ and $w_a$, with typical changes of $\Delta w_0\simeq+0.03$ and $\Delta w_a\simeq-0.06$ for CPL. This reflects the degeneracy between the matter and {DE} densities at late times. The displacement remains subdominant compared to calibration or progenitor-related effects but traces a similar direction in parameter space.
\end{enumerate} Overall, in our analysis, calibration and progenitor evolution are found to be the main drivers of systematic bias in {DE} estimation from SN-Ia cosmology. {Colour scatter, matter-density mismatch, and intergalactic dust produce smaller shifts, given their adopted amplitudes. The $T_\sigma$ metric provides a covariance-weighted measure of these shifts relative to the fiducial model.} 

Across most cases, the parameter shifts follow the $w_0-w_a$ degeneracy direction, suggesting that most systematics primarily distort the inferred $w(z)$. {For the intergalactic-dust template, the sign of the effect is fixed by construction because positive dust attenuation only dims the SN-Ia flux. Therefore, for the adopted dust model, the induced shift in the $w_0-w_a$ plane occurs in a fixed direction. In contrast, calibration offsets, progenitor evolution in $M_B$, progenitor evolution in $x_1$, colour-correction scatter, and density-parameter mismatch} can produce shifts on either side of the degeneracy axis, depending on the sign and magnitude of the injection.

{We also test a second set of injected amplitudes for the CPL and GEN models, shown in Figs.~\ref{fig:cpl_old_vs_new} and \ref{fig:gen_old_vs_new}, respectively. Varying the amplitudes changes the magnitude of the induced parameter shifts, but the qualitative hierarchy remains the same: calibration and progenitor-related effects give the largest biases, while GEN remains less sensitive than CPL.} {These tests also clarify the distinction between shifts of the contour centres and changes in contour area.}

{For the coherent fixed-injection cases, systematics change the mock $\Delta\mu_{\rm SN}$, while the covariance matrix is kept fixed. Therefore, these systematics mainly displace the contour centres rather than substantially increasing the contour area. The colour-scatter case behaves differently because it} {is implemented as an additional diagonal noise term in the SN-Ia covariance matrix rather than as a coherent shift in the mock distance moduli. Varying its amplitude, therefore mainly changes the contour width, with only small changes in the recovered best-fit parameters. More general changes in contour size and orientation are discussed separately in the real-data nuisance-parameter analysis in Section~\ref{sec:realsys}.}

\subsection{Comparative Analysis of DE Models}

\begin{figure}
    \centering
    \includegraphics[width=\linewidth]{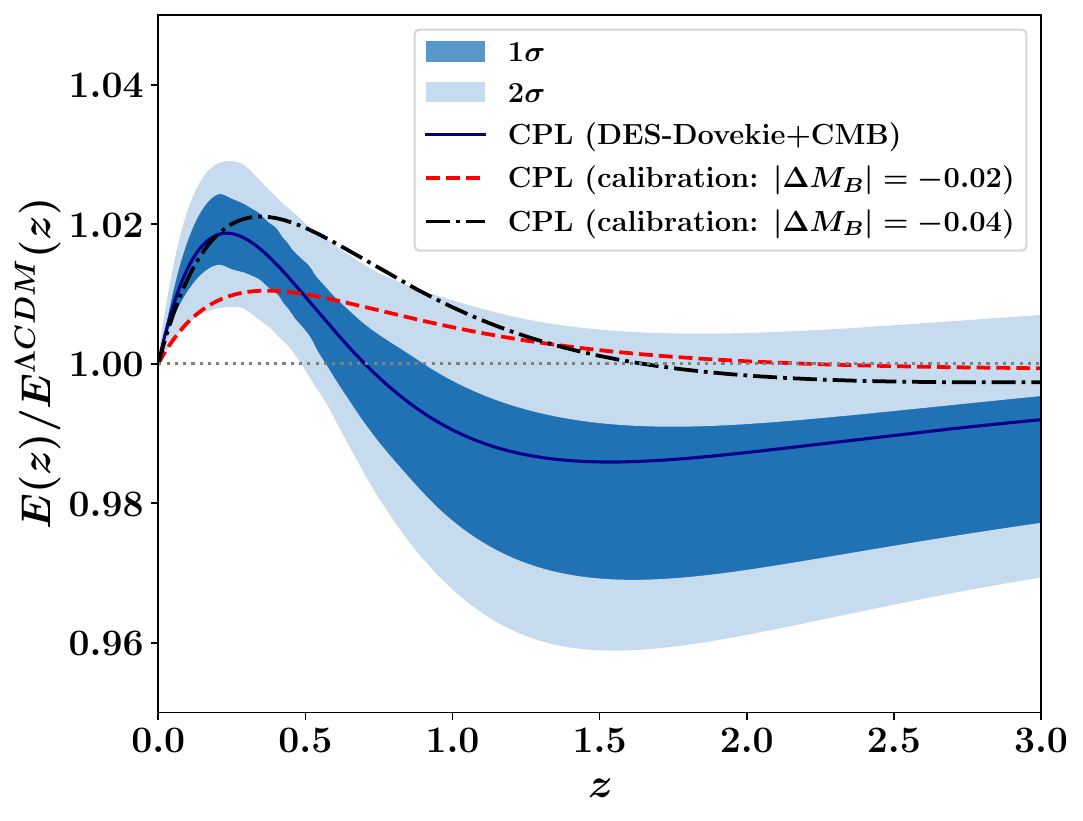}
    \caption{Reconstructed $E(z)/E_{\Lambda{\rm CDM}}(z)$ for the CPL parametrization. The solid curve with shaded bands represents the DESDovekie+compressed CMB constraint, while the dashed and dash-dotted curves show the effect of calibration-systematic injections with $\Delta M_B=-0.02$ and $-0.04$.}
    \label{fig:Ez_real_vs_systematics}
\end{figure}

\begin{figure*}
    \centering
    \includegraphics[width=\linewidth]{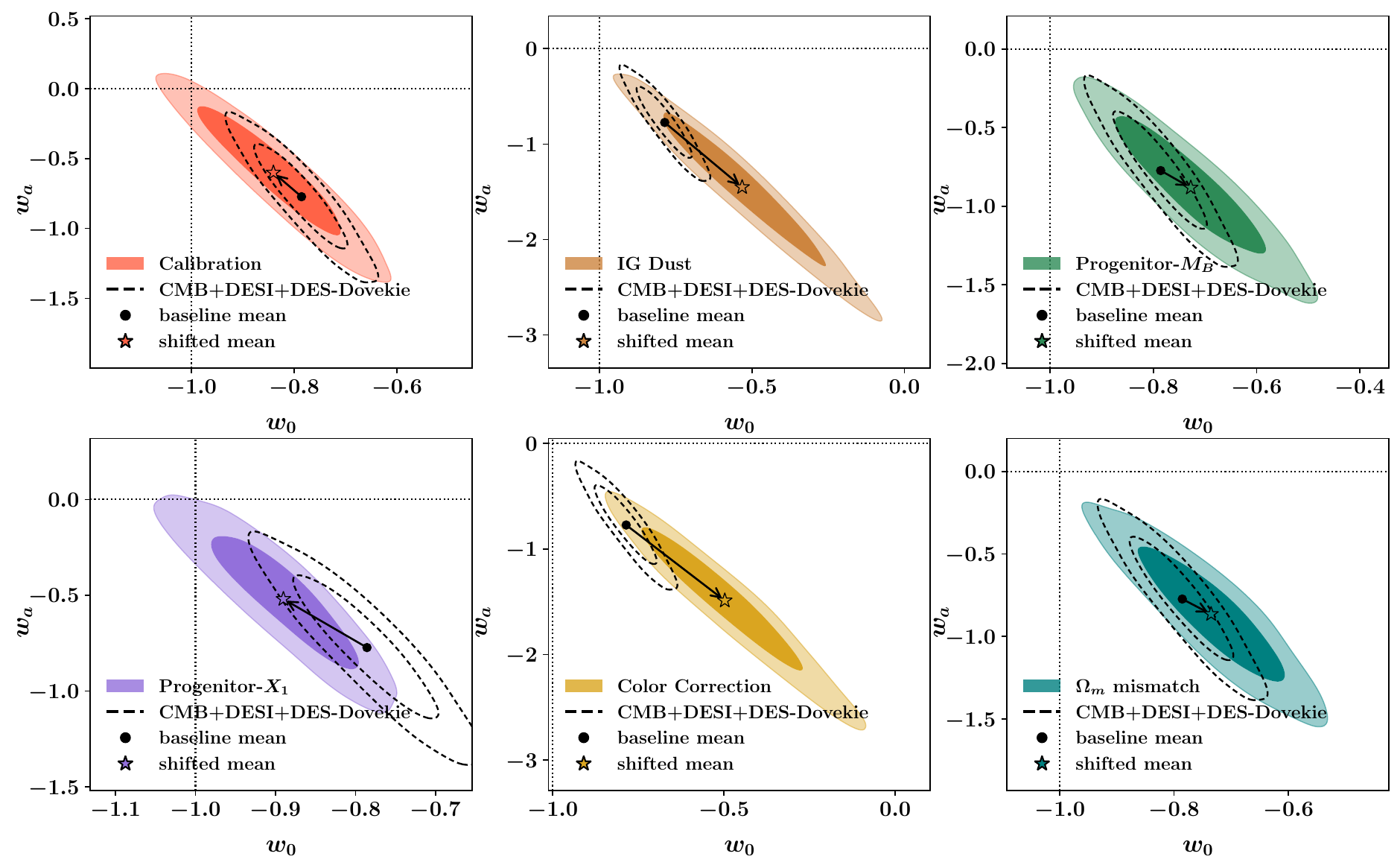}
    \includegraphics[width=0.75\linewidth]{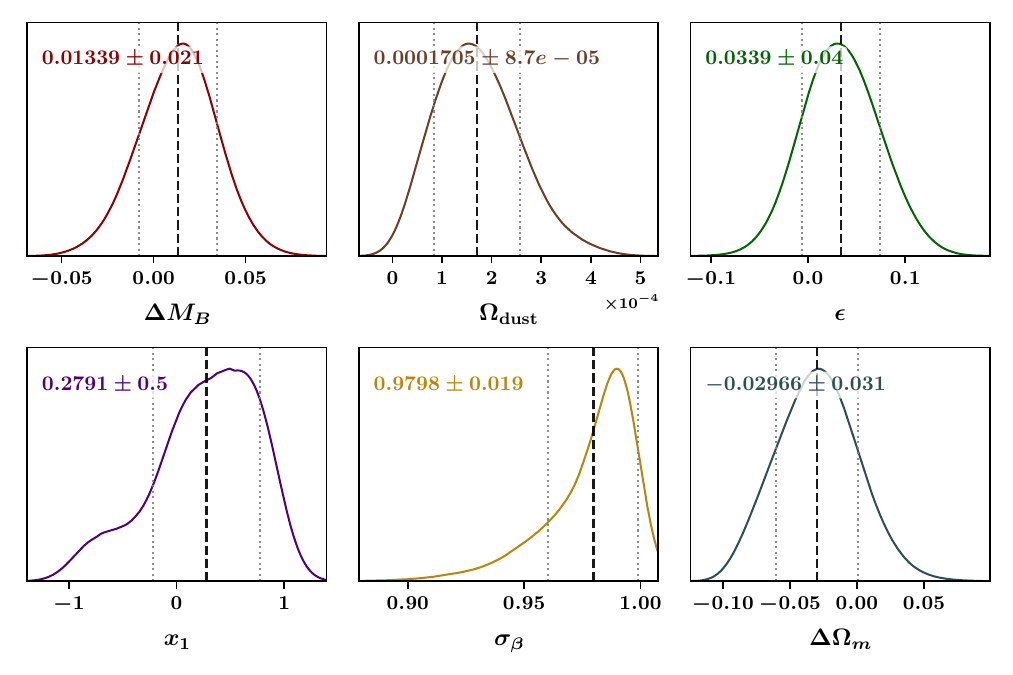}
\caption{{Top panels: shifts in the recovered $w_0-w_a$ constraints for the CPL parametrization when the systematic terms are treated as nuisance parameters and marginalized over. The dashed black contour denotes the baseline CMB+DESI+DES-Dovekie result, and the filled coloured contour shows the corresponding shifted constraint. Bottom panels: 1D posteriors of the nuisance parameters.}    \label{fig:systematics}}
\end{figure*}

The impact of systematics is not uniform across different {DE} parametrizations. The intrinsic flexibility and degeneracy structure of each model determine how it responds to observational biases, influencing both the amplitude and direction of shifts in the $w_0-w_a$ plane. A comparative analysis with respect to the commonly used CPL reveals trends in model sensitivity and stability: \begin{enumerate}[left=0pt]
    \item The JBP model exhibits the largest systematic-induced displacements, particularly under calibration of $M_B$ and progenitor-$x_1$ evolution. Its quadratic dependence on $z/(1+z)$ makes JBP more sensitive to redshift-dependent variations in luminosity, amplifying small biases into stronger shifts in the $w_0-w_a$ plane compared to the linear CPL form.
    \item The LOG parametrization closely mirrors the CPL scenario in both direction and magnitude of systematic shifts. Deviations arise only marginally, mainly for intergalactic dust, suggesting that a logarithmic evolution in $w(z)$ does not provide significant improvement in robustness over the linear CPL form.  
    \item The GEN model shows the least sensitivity to all systematic effects, particularly those arising from calibration and progenitor evolution. Its simpler functional form reduces sensitivity to redshift-dependent biases, making it the most robust among the tested models.  This arises because, unlike CPL, JBP, and LOG, which directly parametrize $w(z)$, the GEN model parametrizes the expansion rate $E(z)$ (or equivalently, the energy density). Since $E(z)$ connects to distance measures through a single integration, whereas $w(z)$ affects them through a double integral, it appears that the GEN model is less affected by cumulative biases in the observed SN-Ia data.
\end{enumerate}
Overall, these comparisons establish a clear trend in how SN-Ia systematics affect {DE} inference: JBP is the most affected, CPL and LOG show intermediate sensitivity, while GEN remains the least affected. This hierarchy demonstrates that the choice of the model or parametrization governs how strongly the inferred values of {DE EoS} parameters, $w_0$ and $w_a$, are susceptible to biases from supernova systematics. Since current SN-Ia data mainly probe $z \lesssim 1$, model selection plays a critical role in interpreting these constraints reliably. 

{The response to a given systematic depends on both its injected amplitude and its redshift dependence. A systematic whose redshift dependence closely follows the distance-redshift behaviour allowed by a given DE parametrization can produce a larger apparent shift in $(w_0,w_a)$ than another systematic with a less degenerate redshift structure. We therefore compare the reconstructed $E(z)$ and $w(z)$ curves in addition to the parameter shifts, since these quantities show how each systematic maps onto the cosmological evolution allowed by each model.} 

Fig.~\ref{fig:residual} shows the impact of the injected systematics on the reconstructed cosmological quantities---the normalised Hubble parameter $E(z)$ (left panels) and the {DE EoS} $w(z)$ (right panels). The upper panels correspond to the CPL model, while the lower panels show the results for the GEN parameterisation. Each curve represents a different source of systematic effect, including calibration offset, color correction, progenitor mass and metallicity evolution, intergalactic dust, and matter density mismatch, compared to the fiducial (dotted black) cosmology. We find that all sources of systematics induce redshift-dependent deviations in the reconstructed quantities, with the calibration and progenitor effects producing the largest biases at low redshifts. 

For the CPL case, the deviations in $E(z)$ rise rapidly at low $z<1$, reaching a maximum around $z \approx 0.5$, and gradually decline towards higher $z>1$. The corresponding impact on the reconstructed DE EoS  $w(z)$ manifests as a smooth phantom-to-non-phantom transition, effectively mimicking a mild apparent evolution of $w(z)$ with redshift. In contrast, for the GEN parametrisation, the deviations in $E(z)$ begin to emerge at low redshift, gradually increasing and reaching a maximum around $z \sim 1$, before slowly declining and settling to an approximately constant offset at higher $z$. For $w(z)$, the behaviour differs qualitatively from the CPL case: the reconstructed EoS tends to remain either entirely in the phantom or non-phantom regime across the redshift range, with systematics primarily shifting its overall amplitude rather than inducing transitions. This pattern underscores that while the CPL model captures a specific range of DE behaviours, the GEN parametrization can accommodate for a broader spectrum, with reconstructed quantities exhibiting distinct responses to underlying uncertainties.

{Figure~\ref{fig:Ez_real_vs_systematics} compares the CPL reconstruction from the DES-Dovekie SN-Ia sample combined with the compressed CMB likelihood with illustrative calibration-like systematic injections. The solid blue curve shows the best-fit real-data reconstruction of $E(z)/E_{\Lambda{\rm CDM}}(z)$, with the shaded regions denoting the $1\sigma$ and $2\sigma$ uncertainties. The real-data best fit shows a low-redshift enhancement around $z\sim0.3$--$0.55$, and the calibration-injection templates with $\Delta M_B=-0.02$ and $-0.04$ produce a qualitatively similar bump-like feature. Other coherent redshift-dependent SN-Ia systematics, such as progenitor evolution in $x_1$, can also generate similar trends. At higher redshifts, the real-data best fit instead shows a dip below the $\Lambda$CDM expectation, with a broad minimum around $z\sim1$--$2$. Although the simple calibration templates do not reproduce the high-$z$ best-fit dip, they remain within the reconstructed $2\sigma$ CL for $z>0.7$. Thus, coherent SN-Ia systematics can mimic part of the reconstructed low-redshift structure and remain broadly consistent with the current uncertainties, but do not reproduce the full best-fit redshift dependence of the real-data reconstruction.}

\subsection{Impact of systematics on real data} \label{sec:realsys}

\begin{table}
\centering
\caption{{Minimum $\chi^2$ and $\Delta\chi^2_{\rm min}=\chi^2_{\rm sys}-\chi^2_{\rm no\,sys}$ values for the baseline CPL and systematic-extended real-data fits.}}
\renewcommand{\arraystretch}{1.2}
\setlength{\tabcolsep}{6pt}
\begin{tabular}{lcc}
\toprule
\textbf{Model / Systematic} & $\boldsymbol{\chi^2_{\rm min}}$ & $\boldsymbol{\Delta\chi^2_{\rm min}}$ \\
\midrule
Baseline, no systematic & $1643.79$ & $0$ \\
\midrule
Calibration offset, $\Delta M_B$ & $1643.43$ & $-0.35$ \\
Intergalactic dust, $\Omega_{\rm dust}$ & $1641.07$ & $-2.71$ \\
Progenitor evolution in $M_B$, $\epsilon$ & $1643.28$ & $-0.51$ \\
Progenitor evolution in $x_1$ & $1643.60$ & $-0.19$ \\
Density parameter mismatch, $\Delta\Omega_m$ & $1642.92$ & $-0.87$ \\
\bottomrule
\end{tabular}
\label{tab:chi2_systematics}
\end{table}

{This section tests how the systematic templates affect the real-data constraints when their amplitudes are allowed to vary. We use the DES-Dovekie SN-Ia sample together with DESI BAO and the compressed CMB likelihood, and restrict this test to the CPL parametrization. Each systematic amplitude is treated as a nuisance parameter and sampled jointly with the cosmological parameters in the MCMC analysis.}

{Figure~\ref{fig:systematics} summarizes the real-data analysis with systematic amplitudes treated as nuisance parameters. The upper panels show the constraints in the $w_0-w_a$ plane. The dashed black contours correspond to the baseline CMB+DESI+DES-Dovekie result. The filled coloured contours show the constraints obtained after marginalizing over each systematic term. The black dot and star mark the posterior means before and after including the systematic, respectively. The lower panels show the one-dimensional posteriors of the nuisance parameters, but should not be interpreted as detections of individual systematics. Rather, they show that allowing plausible systematic terms to vary can modify the inferred DE constraints, and that the templates we used in the mock-injection study remain relevant when applied to current SN-Ia data. We also report the corresponding $\chi^2_{\rm min}$ and $\Delta \chi^2_{\rm min}$ values in Table~\ref{tab:chi2_systematics}, where $\Delta \chi^2_{\rm min}=\chi^2_{\rm sys}-\chi^2_{\rm no\,sys}$, to compare the baseline and systematic-extended fits.}

{The nuisance parameters modify the real-data constraints in different ways. Calibration, progenitor-$M_B$, progenitor-$x_1$, and $\Omega_m$ mismatch mainly shift the posterior mean relative to the baseline result, while the shifted contours still overlap substantially with the baseline constraint. The calibration and progenitor-$x_1$ cases move the posterior mean closer to the $\Lambda$CDM point $(-1,0)$, whereas progenitor-$M_B$ and $\Omega_m$ mismatch move the mean away from this point, with a mild increase in the contour area. In contrast, intergalactic dust and color correction produce major changes in the size and orientation of the contours, indicating a reshaping of the allowed parameter space rather than only a displacement of the posterior mean. For these cases, the impact is not simply a shift of the best-fit point, but a noticeable change in the extent and orientation of the allowed region.}

{We exclude the colour-scatter case from the $\chi^2_{\rm min}$ comparison because it enters as an additional diagonal contribution to the SN-Ia covariance, rather than as a coherent shift in the distance moduli. Since its amplitude is varied freely over the range $[0,1]$, the fit can lower $\chi^2_{\rm min}$ simply by inflating the SN-Ia uncertainties, which would not correspond to a physical improvement of the fit. We therefore do not include a separate colour-scatter entry in Table~\ref{tab:chi2_systematics}.}

\section{Summary and Discussions \label{sec:summary}}

In this work, we have examined how residual SN-Ia systematics propagate into constraints on dynamical DE using both controlled mock injections and a complementary real-data nuisance-parameter analysis. For the representative amplitudes considered here, calibration- and progenitor-related effects produce the largest shifts, while the response to a given systematic depends strongly on the adopted DE parametrization. The key implication is therefore not that one parametrization is intrinsically more reliable than another, but that the same systematic distortion can project very differently onto different functional forms. For a fixed amplitude and redshift dependence, the displacement in the $(w_0,w_a)$ plane can differ substantially in both magnitude and direction. The inferred evidence for DE evolution is thus shaped not only by the data, but also by the parametrization through which the data are interpreted and by its associated degeneracy structure.

When the late-time expansion history is described by a two-parameter phenomenological form, a coherent distortion of the SN-Ia Hubble diagram is absorbed into the inferred $(w_0,w_a)$ constraints, thereby compressing part of the redshift dependence that could otherwise help distinguish a systematic effect from genuine cosmological evolution. We therefore advocate that SN-Ia constraints on the late-time expansion history be complemented, where possible, by non-parametric or weakly parametric reconstructions of quantities such as $E(z)$ or $\mu(z)$, for example using Gaussian processes, free-form splines, or binned reconstructions with an explicit correlation prior. Although such reconstructions do not remove systematic biases, they preserve the redshift dependence more explicitly. This allows an inferred feature to be compared directly with the systematic templates considered here, rather than appearing only as a displacement in two correlated parameters whose interpretation is itself dependent on the assumed functional form.

From this perspective, phenomenological parametrizations such as $(w_0,w_a)$ are most naturally regarded as compact summary statistics of the reconstructed expansion history rather than, by themselves, as the primary basis for establishing evidence for new physics. Parametrized fitting remains entirely appropriate when the adopted functional form follows from a specified physical model, for example, from a quintessence, modified-gravity, or coupled-DE Lagrangian. In such cases, the functional form is a prediction of the underlying theory and the parameters possess a physical interpretation that is independent of their role in fitting the data. The situation is different for purely phenomenological parametrizations, for which the functional form is imposed as a convenient compression of a much broader space of possible expansion histories. In the absence of a physically motivated functional form, parametrization dependence should itself be part of the robustness assessment, since an apparent preference for evolving DE may partly reflect the chosen parametrization rather than the data alone.

More broadly, our results therefore suggest that possible evidence for new physics in the late-time Universe should be tested primarily at the level of the reconstructed expansion history, with phenomenological parametrizations used as complementary summaries and physically motivated parametrizations used when they constitute genuine predictions of an underlying theory. Our real-data analysis further shows that dominant SN-Ia systematics should be marginalized jointly with the cosmological parameters, since they can modify not only the posterior mean but also the size and orientation of the allowed parameter region.

Finally, the analysis reinforces a central message: although hints of dynamical DE have been reported in combined DESI and CMB analyses without the direct inclusion of SN-Ia data, unaccounted systematics can still bias the inferred late-time expansion history when supernovae are included in joint constraints. The key takeaway for SN-Ia cosmology is therefore continued improvements in calibration and population modelling with explicit systematic marginalization, cross-parametrization consistency tests, and complementary reconstructions of $H(z)$. While improvements in calibration and population modelling are already central to current SN-Ia analyses, the additional guidance from our work is that parametrization dependence itself must also be included in this assessment. A feature that remains stable under explicit systematic marginalization and across distinct phenomenological DE parametrizations, and is subsequently recovered by an independent non-parametric or free-form reconstruction of $H(z)$, would provide substantially stronger evidence for a genuine cosmological departure from $\Lambda$CDM. Conversely, if a particular feature is supported only after projection onto some specific phenomenological form, its interpretation as evidence for new physics should be treated with caution.

\begin{acknowledgments}
PM acknowledges funding from the Anusandhan National Research Foundation (ANRF), Govt of India, under the National Post-Doctoral Fellowship (File no. PDF/2023/001986). AAS acknowledges the funding from ANRF, Govt of India, under the research grant no. CRG/2023/003984. SD acknowledges support from  UK Research and Innovation (UKRI) under the UK government's Horizon Europe funding Guarantee EP/Z000475/1. We acknowledge the use of the HPC facility, Pegasus, at IUCAA, Pune, India. This article/publication is based upon work from COST Action CA21136- ``Addressing observational tensions in cosmology with systematics and fundamental physics (CosmoVerse)'', supported by COST (European Cooperation in Science and Technology).
\end{acknowledgments}




\bibliography{ref}

\end{document}